\documentclass[11pt, a4paper]{article}
\usepackage{styleBuding}
\usepackage{bm}
\usepackage{color}
\usepackage[usenames,dvipsnames,svgnames,table]{xcolor}
\usepackage{amsmath}
\usepackage{amssymb}
\usepackage{graphicx}
\usepackage{slashed}
\usepackage{soul}
\usepackage{multirow}
\usepackage{subfigure}
\usepackage{siunitx} 
\usepackage[utf8]{inputenc}

\pdfoutput=1

\begin{document}

\title{\boldmath The 96~GeV Diphoton Excess
in the Seesaw Extensions of the Natural NMSSM}


\author{Junjie Cao,}
\author{Xinglong Jia,}
\author{Yuanfang Yue,}
\author{Haijing Zhou,}
\author{and Pengxuan Zhu}

\affiliation{Department of Physics, Henan Normal University, Xinxiang 453007, China}

\emailAdd{junjiec@alumni.itp.ac.cn}
\emailAdd{645665975@qq.com}
\emailAdd{yuanfang405@gmail.com}
\emailAdd{zhouhaijing0622@163.com}
\emailAdd{zhupx99@icloud.com}

\abstract{The Next-to Minimal Supersymmetric Standard Model (NMSSM) with a Type-I seesaw mechanism extends the NMSSM  by three generations of right-handed neutrino fields to generate neutrino mass. As a byproduct it renders the lightest sneutrino as a viable DM candidate. Due to the gauge singlet nature of the DM, its scattering with nucleon is suppressed in most cases to coincide spontaneously with the latest XENON-1T results. Consequently, broad  parameter spaces in the Higgs sector, especially a light Higgsino mass, are resurrected as experimentally allowed, which makes the theory well suited to explain the long standing $b \bar{b}$ excess at LEP-II and the continuously observed $\gamma \gamma$ excess by CMS collaboration.  We show by both analytic formulas and numerical results that the theory can naturally predict the central values of the excesses in its broad parameter space, and the explanations are consistent with the Higgs data of the discovered Higgs boson, $B-$physics and DM physics measurements, the electroweak precision data as well as the LHC search for sparticles.  Part of the explanations may be tested by future DM experiments and the SUSY search at the LHC. }

\maketitle

\section{Introduction\label{sec:intro}}

Electroweak symmetry breaking (EWSB) is one of the most important issue in particle physics. The discovery of a Standard Model (SM) like Higgs boson at the Large Hadron Collider (LHC) ~\cite{Aad:1471031,Chatrchyan:1471016} indicates the correctness of Higgs mechanism, while the quadratically divergent correction to the boson  mass in the SM implies the need of a more complex theoretical structure to account for the EWSB in a natural way. Without considering extreme cases, the complete EWSB mechanism should manifest itself by potentially sizable deviations (less than roughly 20\% at 95\% confidence level according to current Higgs data at the LHC~\cite{ATLAS:2019slw}) of the boson's property from its SM prediction, and/or by exotic signals at colliders. Interestingly, so far it seems that the latter has emerged through  a $2.3\sigma$ local excess for the channel $e^+ e^- \to Z \phi_{b \bar{b}} \to Z b \bar b$ at LEP-II  with the scalar mass $m_{\phi_{b \bar{b}}}\sim98~{\rm GeV}$~\cite{Barate:2003sz,Schael:2006cr}, and also through a roughly $3\sigma$ local excess for the channel $p p \to \phi_{\gamma \gamma} \to \gamma \gamma$ at the LHC with $m_{\phi_{\gamma \gamma}} \simeq 96~{\rm GeV}$, which was reported recently by CMS collaboration after combining 8 and 13 TeV data~\cite{Sirunyan:2018aui, CMS-PAS-HIG-14-037}\footnote{We note that the ATLAS collaboration also published its analysis of about $80~{\rm fb}^{-1}$ data on the diphoton signal~\cite{ATLAS:2018xad}, and it claims no significant excess over the SM expectation for the diphoton mass between $65~{\rm GeV}$ and  $110~{\rm GeV}$. This result does not conflict with the CMS observation since the limit of the ATLAS analysis on the signal strength is significantly weaker than the corresponding one obtained by CMS collaboration (see the comparison of the two limits in Fig.~1 of Ref.~\cite{Heinemeyer:2018wzl}). We infer from the two experimental reports that the low performance of the ATLAS analysis is mainly due to its relatively large uncertainties in
determining the diphoton invariant mass $\phi_{\gamma \gamma}$ (see Fig.2 of~\cite{ATLAS:2018xad} and Fig.1 of ~\cite{CMS-PAS-HIG-14-037}) and also in fitting the continuum background of the signal. }. The normalized signal strengths of the excesses are~\cite{Sirunyan:2018aui,Cao:2016uwt}
\begin{eqnarray}
	\mu_{\rm LEP} &=& \frac{\sigma(e^+ e^- \to Z \phi_{b\bar{b}} \to Z b \bar{b})}{\sigma_{\rm SM}(e^+ e^- \to Z H_{\rm SM} \to Z b \bar{b})} = 0.117 \pm 0.057, \\
	\mu_{\rm CMS} &=& \frac{\sigma(p p \to \phi_{\gamma \gamma} \to \gamma\gamma)}{\sigma_{\rm SM} (p p \to H_{\rm SM} \to \gamma\gamma)} = 0.6 \pm 0.2.  \label{mu-Excesses}
\end{eqnarray}

Since the mass resolution for the $b \bar{b}$ final state at LEP-II is rather coarse~\cite{Heinemeyer:2018wzl} and at same time $\phi_{b \bar{b}}$ and $\phi_{\gamma \gamma}$ are so close in mass, it is conjectured that the two excesses do not emerge accidently and may have the same physical origin. So far a variety of  beyond SM theories were studied to reveal it, where the intermediate scalar $\phi$ is usually taken as a gauge singlet CP-even Higgs or a singlet-like particle. These theories include the radion model~\cite{Richard:2017kot,Sachdeva:2019hvk}, the singlet extensions of the SM by additional vector-like matter fields~\cite{Fox:2017uwr,Mondal:2019ldc}, the fermiophobic Type-I Two Higgs Doublet Model (2HDM)~\cite{Fox:2017uwr, Haisch:2017gql}, the singlet extensions of the 2HDM~\cite{Biekotter:2019mib, Biekotter:2019kde} and its supersymmetric version (namely the Next-to-Minimal Supersymmetic Standard Model (NMSSM))~\cite{Cao:2016uwt,Hollik:2018yek,Heinemeyer:2018wzl,Domingo:2018uim,Tao:2018zkx,Beskidt:2017dil,Choi:2019yrv}, as well as the $\mu$-from-$\nu$ Supersymmetric Standard Model ($\mu\nu$SSM)~\cite{Biekotter:2017xmf, Biekotter:2019gtq}. Among these models, the NMSSM~\cite{Ellwanger:2009dp} is of particular interest because, due to the introduction of one singlet Higgs field, it has theoretical advantages over the popular Minimal Supersymmetric Standard Model (MSSM), such as  generating dynamically the $\mu$-term, which is responsible for Higgsino mass,  and providing more feasible dark matter (DM) candidates so that the model's phenomenology is enriched greatly~\cite{Ellwanger:2014hia,Kim:2015dpa,Cao:2016nix,Ellwanger:2018zxt}. Moreover, as far as the light CP-even Higgs scenario of the model (which is appropriate to explain the excesses) is concerned, the mass of the SM-like Higgs boson can be significantly lifted up by both an additional tree-level contribution and the singlet-doublet mixing effect~\cite{Ellwanger:2011aa,Badziak:2013bda,Cao:2012fz}, which makes large radiative corrections from stop loops unnecessary in predicting $m_h \simeq 125~{\rm GeV}$, and the Higgsino mass is upper bounded by about $400~{\rm GeV}$, which is the right range to predict $Z$ boson mass in a natural way~\cite{Cao:2012fz,Cao:2013gba,Jeong:2014xaa}.

A thorough analysis of the light Higgs scenario in the general NMSSM was recently performed in~\cite{Choi:2019yrv} by both compact analytic formulas and numerical results.  It was concluded that there are parameter spaces where the excesses can be well explained without conflicting with the $125~{\rm GeV}$ Higgs data collected at the LHC.  With regard to such a study, we remind that, for the NMSSM as one of the most intensively studied supersymmetric theory, its parameter space in Higgs sector has been tightly limited by DM physics and also by the LHC search for sparticles~\cite{Cao:2018rix}, which should be considered in a comprehensive study of the excesses. The tightness of the constraints comes from following facts:
\begin{itemize}
\item In the NMSSM, the lightest neutralino (usually with Bino or Singlino field as its dominant component) is customarily taken as a DM candidate. Some parameters in the neutralino sector of the theory, such as $\lambda$, $\kappa$, $\tan \beta$ and $\mu$, are also inputs of the Higgs sector~\cite{Ellwanger:2009dp}. Notably,
    besides affecting the Higgs mass spectrum, these parameters usually play an important role in determining the DM property, such as its mass, field component as well as interactions with Higgs and SM gauge bosons~\cite{Cao:2015loa}. So they are restricted by the measurements in DM physics.
\item Given that squarks and sleptons are preferred heavy by the direct search for sparticles at the LHC and consequently they have little effect on DM physics when $m_{\rm DM} \sim 100~{\rm GeV}$,  the Higgs bosons and the SM gauge bosons often act as the mediators or final states of DM annihilation. Then the DM relic density precisely measured by WMAP and Planck experiments~\cite{Aghanim:2018eyx} requires fine-tuned configurations of the involved parameters~\cite{Cao:2014efa,Cao:2015loa}.
\item So far XENON-1T experiment has reached unprecedent sensitivity in detecting DM-nucleon scattering, i.e. $\sigma^{\rm SI} \sim 10^{-47}~{\rm cm^2}$ for spin-independent (SI) cross section~\cite{Aprile:2018dbl} and  $\sigma^{\rm SD} \sim 10^{-41}~{\rm cm^2}$ for spin-dependent (SD) cross section~\cite{Aprile:2019dbj}.  Since the t-channel exchange of Higgs bosons (Z boson) is the dominant contribution to the SI (SD) cross section at tree level,  the experiment can exclude a large portion of the parameter space in the Higgs sector, especially in case of light Higgsinos and/or light Higgs bosons where strong cancellation between different Higgs contributions must be present to coincide with the experimental results~\cite{Cao:2016cnv,Badziak:2015exr}.
\item With the smooth progress of the LHC in looking for electroweakinos by multi-lepton signals, the mass spectrum of neutralinos and charginos has been tightly limited within certain patterns for $\mu \lesssim 500~{\rm GeV}$~\cite{Cao:2018rix}.
\end{itemize}
In fact, we once studied the excesses in the NMSSM with a $\mathbb{Z}_3$ discrete symmetry by considering the constraints from LUX and PandaX experiments on the DM-nucleon
scattering in 2016. We found that they can be explained at $1\sigma$ level only in a very narrow parameter space and at the cost of relaxing the relic density constraint~\cite{Cao:2016uwt}. Since the latest XENON-1T results have improved the previous sensitivity by a factor of about 5, we checked that the space becomes experimentally disfavored~\cite{Cao:2018rix}.

Given the great theoretical advantages of the NMSSM and unfortunately the strong experimental constraints on its most natural parameter space, we were motivated to augment the NMSSM with different seesaw mechanisms to generate neutrino mass and also to enable the lightest sneutrino $\tilde{\nu}_1$ as a viable DM candidate~\cite{Cao:2017cjf,Cao:2018iyk,Cao:2019aam,Cao:2019qng}. The general feature of such extensions is that the singlet Higgs field plays extra roles~\cite{Cao:2017cjf,Cao:2018iyk}: apart from being responsible for heavy neutrino mass via the interaction $\hat{S} \hat{\nu} \hat{\nu}$ in superpotential ($\hat{S}$ and $\hat{\nu}$ denote the superfields of the singlet Higgs and the heavy neutrino respectively), it contributes to the annihilation of $\tilde{\nu}_1$ and consequently makes the property of $\tilde{\nu}_1$ compatible with various measurements in a natural way. This can be understood from two popular cases. One is that the singlet Higgs can mediate the transition between the $\tilde{\nu}_1$ pair and the Higgsino pair so that $\tilde{\nu}_1$ and  the Higgsinos are in thermal equilibrium in early universe before their freeze-out. If their mass splitting is less than about $10\%$, the number density of the Higgsinos can track that of $\tilde{\nu}_1$ during freeze-out, and as a result the Higgsinos played a dominant role in determining the density due to its relatively strong interaction with SM particles~\cite{Coannihilation} (in literature such a phenomenon was called coannihilation \cite{Griest:1990kh}). In this case, the constraint on the Higgsino mass $\mu$ from the LHC search for electroweakinos is rather weak  due to the compressed spectrum, and light Higgsinos with $\mu \sim 100~{\rm GeV}$ are still allowed. The other is that, due to its gauge singlet nature,  $\tilde{\nu}_1$ and the singlet Higgs can compose a secluded DM sector where the measured relic abundance can be accounted for by the annihilation of $\tilde{\nu}_1$ into a pair of the singlet Higgs. In both the cases, $\tilde{\nu}_1$ couples very weakly with the SM particles so that its scattering with nucleon is always suppressed, which is consistent with current DM direct detection (DD) results. This is a great theoretical advantage in light of the tight experimental limit. At this stage, we emphasize that, when one fixes the mass spectrum of the Higgs bosons and neutralinos, it usually happens that the theory is kept compatible with various DM measurements only by adjusting the parameters in the sneutrino sector~\cite{Cao:2017cjf,Cao:2018iyk,Cao:2019aam}. This reflects the fact that, although the DM sector and the Higgs sector of the theory are entangled together to survive various experimental constraints, which is the same as the NMSSM, their correlation becomes loose and the constraints from DM physics are weakened greatly. This will resurrect broad  parameter spaces in the Higgs sector as experimentally allowed and thus make the theory suitable to explain the excesses. This feature was not noticed in the previous works~\cite{Cao:2017cjf,Cao:2018iyk,Cao:2019aam}, and studying the capability of the augmented theory to explain the excesses is the main aim of this work.

This paper is organized as follows. In section \ref{Section-Model}, we first take the NMSSM with a Type-I seesaw mechanism (Type-I NMSSM) as an example to recapitulate the basics of the more general framework where the NMSSM is augmented by different seesaw mechanisms,  including its field content and Lagrangian, then we
turn to discuss the conditions to produce sizable $b\bar{b}$ and $\gamma \gamma$ signals. In Section \ref{Section-excess},  we perform a comprehensive scan over the vast parameter space of the Higgs sector to look for the regions where the excesses can be well explained. In this process, we consider some experimental results, such as the Higgs data of the discovered Higgs, $B-$physics measurements as well as precision electroweak measurements, to limit the parameter space, and plot the map of the profile likelihood (PL) for the excesses on different planes to study their implication in the theory. In Section \ref{Section-constraint}, we further study the constraints from the DM physics and the sparticle search, and point out that some explanations can easily survive the constraints. We also choose one parameter point to show that the fine tunings associated with the excesses are not serious. Conclusions are made in Section \ref{Section-Conclusion}.

\begin{table*}[t]
	\begin{center}
		\begin{tabular}{|c|c|c|c|c|c|}
			\hline \hline
			Superfield & Spin 0 & Spin \(\frac{1}{2}\) & Generations & \((U(1)\otimes\, \text{SU}(2)\otimes\, \text{SU}(3))\) \\
			\hline
			\(\hat{q}\) & \(\tilde{q}\) & \(q\) & 3 & \((\frac{1}{6},{\bf 2},{\bf 3}) \) \\
			\(\hat{l}\) & \(\tilde{l}\) & \(l\) & 3 & \((-\frac{1}{2},{\bf 2},{\bf 1}) \) \\
			\(\hat{H}_d\) & \(H_d\) & \(\tilde{H}_d\) & 1 & \((-\frac{1}{2},{\bf 2},{\bf 1}) \) \\
			\(\hat{H}_u\) & \(H_u\) & \(\tilde{H}_u\) & 1 & \((\frac{1}{2},{\bf 2},{\bf 1}) \) \\
			\(\hat{d}\) & \(\tilde{d}_R^*\) & \(d_R^*\) & 3 & \((\frac{1}{3},{\bf 1},{\bf \overline{3}}) \) \\
			\(\hat{u}\) & \(\tilde{u}_R^*\) & \(u_R^*\) & 3 & \((-\frac{2}{3},{\bf 1},{\bf \overline{3}}) \) \\
			\(\hat{e}\) & \(\tilde{e}_R^*\) & \(e_R^*\) & 3 & \((1,{\bf 1},{\bf 1}) \) \\
			\(\hat{s}\) & \(S\) & \(\tilde{S}\) & 1 & \((0,{\bf 1},{\bf 1}) \) \\
			\(\hat{\nu}\) & \(\tilde{\nu}_R^*\) & \(\nu_R^*\) & 3 & \((0,{\bf 1},{\bf 1}) \) \\
			\hline \hline
		\end{tabular}
	\end{center}
\vspace{-0.4cm}

\caption{Field content of the NMSSM with Type-I seesaw mechanism.}
\label{table1}
\end{table*}

\section{Theoretical preliminaries}  \label{Section-Model}

\subsection{NMSSM with the Type-I seesaw mechanism}

As the simplest extension of the NMSSM, the Type-I NMSSM augments the NMSSM by three generation right-handed neutrino fields to generate neutrino masses. With the field content presented in Table \ref{table1}, its superpotential $W$ and soft breaking terms $L_{\rm soft}$  are~\cite{Cerdeno:2008ep,Cerdeno:2009dv}
\begin{eqnarray}
W &=& W_F+\lambda \hat{s} \hat{H}_u \cdot  \hat{H}_d
+\frac{1}{3} \kappa \hat{s}^3+\bar{\lambda}_{\nu} \hat{s} \hat{\nu} \hat{\nu} +Y_\nu \,\hat{l} \cdot \hat{H}_u \,\hat{\nu},  \label{Superpotential} \nonumber \\
L_{\rm soft} &=&  m_{H_d}^2 |H_d|^2 +m_{H_u}^2 |H_u|^2 +m_S^2 |S|^2 + \bar{m}_{\tilde\nu}^{2} \tilde{\nu}_{R}\tilde{\nu}^*_{R}
\nonumber \\
&& + ( \lambda A_\lambda S H_u\cdot H_d  + \frac{1}{3} \kappa A_\kappa S^3 + \bar{\lambda}_{\nu} \bar{A}_{\lambda_{\nu}}S \tilde{\nu}^*_{R} \tilde{\nu}^*_{R}
+ Y_\nu \bar{A}_{\nu} \tilde{\nu}^*_{R} \tilde{l} H_u + \mbox{h.c.})  + \cdots
\label{superpotential}
\end{eqnarray}
where $W_F$ denotes the superpotential of the MSSM without the $\mu$ term, and a $\mathbb{Z}_3$ symmetry is considered to forbid the appearance of any dimensional parameters in $W$. The coefficients $\lambda$ and $\kappa$ parameterize the interactions among the Higgs fields, and $Y_\nu$ and $\bar{\lambda}_\nu$ are neutrino Yukawa couplings with flavor index omitted to make the formulas concise and more intuitive.
Since the soft breaking squared masses $m_{H_u}^2$, $m_{H_d}^2$ and $m_S^2$ are related with the vacuum expectation values of the fields $H_{u}$, $H_d$ and $S$,
$ \langle H_u \rangle = v_u/\sqrt{2}$, $ \langle H_d \rangle = v_d/\sqrt{2}$ and $\langle S \rangle = v_s/\sqrt{2}$,
by the minimization conditions of the Higgs potential after the electroweak symmetry breaking~\cite{Ellwanger:2009dp}, it is customary to
take $\lambda$, $\kappa$, $\tan \beta \equiv v_u/v_d$, $A_\lambda$, $A_\kappa$ and $\mu \equiv \lambda v_s/\sqrt{2} $ as theoretical input parameters of the Higgs sector.

Same as the NMSSM, one usually introduces following combinations of the Higgs fields:
\begin{eqnarray}
H_1=\cos{\beta} H_u+\varepsilon \sin{\beta} H_d^* ,~~H_2=\sin{\beta} H_u - \varepsilon \cos{\beta} H_d^*,~~H_3=S
\end{eqnarray}
where $\varepsilon$ is two-dimensional antisymmetric tensor, i.e. $\varepsilon_{12}=-\varepsilon_{21}=1$ and $\varepsilon_{11}=\varepsilon_{22}=0$. In this representation, $H_i$ ($i=1,2,3$) take the following form:
    \begin{eqnarray}
        H_1 =   \left ( \begin{array}{c}
                    H^+ \\
                    \frac{S_1 + \mathrm{i} P_1}{\sqrt{2}}
                \end{array} \right),~~
        H_2 &=& \left ( \begin{array}{c}
                    G^+ \\
                    v + \frac{S_2 + \mathrm{i} G^0}{\sqrt{2}}
                \end{array} \right),~~
        H_3  =  v_s +\frac{1}{\sqrt{2}} \left( S_3 + \mathrm{i} P_2 \right),
    \end{eqnarray}
These expressions indicate that the field $H_2$ corresponds to the SM Higgs field, and the fields $S_1$ , $S_2$ and $S_3$  mix to form three physical CP-even Higgs bosons.
Therefore, the CP-even Higgs boson with largest $S_2$ component is called the SM-like Higgs boson. In the
basis $(S_1,~ S_2, ~S_3)$, the mass matrix is given by~\cite{Cao:2012fz}:
\begin{equation}\label{CP-even Mass}
\begin{split}
    M_{11}^2&=\frac{2\mu(\lambda A_{\lambda}+\kappa \mu)}{\lambda {\sin} 2\beta}+ (m_Z^2-\frac{1}{2} \lambda^2 v^2){\sin}^2 2\beta \\
    M_{12}^2&=-\frac{1}{4}(2m_Z^2-\lambda^2 v^2){\sin} 4\beta \\
    M_{13}^2&=-\sqrt{2}(\lambda A_{\lambda}+2\kappa \mu)v {\cos} 2\beta \\
    M_{22}^2&=m_Z^2{\cos}^2 2\beta+ \frac{1}{2} \lambda^2 v^2 {\sin}^2 2\beta \\
    M_{23}^2&=\frac{v}{\sqrt{2}} [2\lambda \mu -(\lambda A_{\lambda}+2\kappa \mu){\sin}2\beta] \\
    M_{33}^2&=\frac{\lambda^2 v^2 A_{\lambda}{\sin}2\beta}{4\mu}+\frac{\mu}{\lambda}(\kappa A_{\kappa}+4\kappa^2 \frac{\mu}{\lambda})
\end{split}
\end{equation}
where the expression of $M_{22}^2$ indicates that the SM Higgs mass at tree level gets an additional contribution  $\frac{1}{2} \lambda^2v^2 \sin^2 2 \beta$ in comparison with corresponding MSSM prediction. The matrix also indicates
that if the relation $M_{33}^2 < m_{22}^2$ holds, the mixing between the fields $S_2$ and $S_3$ can further
enhance the mass of the $S_2$-dominated state. Benefiting from the contributions, the SM-like Higgs boson
does not need a large radiative contribution from stop loops to get its mass around $125~ {\rm GeV}$ ~\cite{Ellwanger:2011aa,Cao:2012fz,Badziak:2013bda}.
Due to the attractive feature, this case was called natural NMSSM in literature~\cite{King:2012tr}.
The model also predicts two CP-odd mass eigenstates $A_i$ ($i=1,2$), which are the mixtures of the fields $P_1$ and $P_2$, and a pair of charged Higgs bosons $H^\pm = \cos \beta H_u^\pm + \sin \beta H_d^\pm$.
Throughout this paper, we label the neutral eigenstates in an ascending mass order, i.e.
$m_{h_1} < m_{h_2} < m_{h_3}$ and $m_{A_1} < m_{A_2}$.

The Higgs sector of the model has following features:
\begin{itemize}
\item One CP-even state corresponds to the Higgs boson discovered at the LHC. Since experimental measurements require its property quite SM Higgs like,
the mixing of the $S_2$ field with the other fields should be less than about $10\%$~\cite{ATLAS:2019slw}. This implies from the definition of the $S_1$ and $S_2$ fields that it is
$ {\rm Re}[H_u^0]$ dominated if $\tan \beta \gg 1$, and the heavy doublet-dominated state is mainly composed by ${\rm Re}[H_d^0]$.
\item Similar to the situation of the MSSM, the heavy doublet-dominated CP-even state is roughly degenerate in mass with the doublet-dominated CP-odd state, and also with the charged states. The LHC search for extra Higgs bosons together with the indirect
constraints from $B$-physics have required $m_{H^\pm} \gtrsim 0.5{\rm TeV}$~\cite{Bagnaschi:2017tru}.
\item With regard to the singlet-dominated states, they may be very light without conflicting with relevant collider constraints. One new function of these states is that they can couple directly with the sneutrino pair by three and four scalar interactions, which are induced by the $\bar{\lambda}_{\nu}\,\hat{s}\,\hat{\nu}\,\hat{\nu}$ term in the superpotential and its soft breaking term. As a result, they may appear as the final state of the sneutrino pair annihilation in early universe or mediate the annihilation,  and thus play an important role in sneutrino DM physics.
\end{itemize}

In the following, we recapitulate the features of the neutrino and sneutrino sectors in the Type-I NMSSM, which differ greatly from those of the NMSSM. We first
focus on the neutrino sector. The neutrino Yukawa interactions take following form
\begin{eqnarray}
{\cal L}_{\nu} = \nu_R^\ast Y_\nu H_u^0 \nu_L   + \nu_R^\ast \bar{\lambda}_\nu S \nu_R^\ast + {\rm {~h.c.}},
\end{eqnarray}
and they are responsible for neutrino masses after the involved Higgs fields develop vevs. In the interaction basis $(\nu_L, \nu_R^\ast)$,
the $6\times 6$ neutrino mass matrix reads
\begin{eqnarray}
M_{\rm{Type-I}}=\left(\begin{array}{c c} 0 & \frac{v_u}{\sqrt{2}} Y_\nu \\ \frac{v_u}{\sqrt{2}} Y_\nu^T & \sqrt{2} v_s \bar{\lambda}_\nu \end{array}\right)\,,  \label{Neutrino-mass}
\end{eqnarray}
and given that the magnitude of the right-handed neutrino mass matrix $M = \sqrt{2} v_s \bar{\lambda}_\nu$ is much larger than that of $\frac{v_u}{\sqrt{2}} Y_\nu$, the heavy fields can be integrated out to get the $3\times3$ mass matrix of light active neutrinos~\cite{Kitano:1999qb}, $ M_{\nu} = \frac{1}{2}Y_{\nu} v_u M^{-1} Y_{\nu}^T v_u$.
This symmetric effective mass matrix can be diagonalized by the unitary Pontecorvo-Maki-Nakagawa-Sakata (PMNS)
matrix as follows
\begin{eqnarray}
 U_{\rm PMNS}^T M_{\mathrm{\nu}} U_{\rm PMNS} = \mathrm{diag}(m_{\nu_1}\,, m_{\nu_2}\,, m_{\nu_3})\,,
\end{eqnarray}
with $m_{\nu_1}$, $m_{\nu_2}$ and $m_{\nu_3}$ denoting the masses of the three lightest neutrinos. Since the PMNS matrix has been
determined by neutrino experiments (especially by neutrino oscillation data)~\cite{PDG}, one can express the Yukawa coupling matrix $Y_\nu$ in terms
of the $U_{\rm PMNS}$ by a modified Casas-Ibarra parameterization \cite{Casas:2001sr},
\begin{eqnarray}
 \frac{v_u}{\sqrt{2}} Y_\nu^T = V^\dagger \mathrm{diag}(\sqrt{M_1}\,,\sqrt{M_2}\,,\sqrt{M_3})\; R\; \mathrm{diag}(\sqrt{m_{\nu_1}}\,, \sqrt{m_{\nu_2}}\,, \sqrt{m_{\nu_3}}) U^\dagger_{\rm PMNS}\,.
\end{eqnarray}
where $V$ is a unitary matrix that diagonalizes $M$ by $M=V^\dagger \mathrm{diag}(M_1\,, M_2\,, M_3) V^*$,
and $R$ is a complex orthogonal matrix given by
\begin{eqnarray}
R = \left( \begin{array}{ccc} c_{2} c_{3}
& -c_{1} s_{3}-s_1 s_2 c_3& s_{1} s_3- c_1 s_2 c_3\\ c_{2} s_{3} & c_{1} c_{3}-s_{1}s_{2}s_{3} & -s_{1}c_{3}-c_1 s_2 s_3 \\ s_{2}  & s_{1} c_{2} & c_{1}c_{2}\end{array} \right) \,,
\end{eqnarray}
with $c_i\equiv \cos \theta_i$, $s_i\equiv \sin\theta_i$ and
$\theta_1$, $\theta_2$, and $\theta_3$ being arbitrary angles. This formula shows that the neutrino Yukawa coupling $Y_\nu$ is generally flavor non-diagonal,
and for $m_{\nu_i} \sim  0.1~{\rm eV}$ indicated by neutrino experiments and $M_i \sim {\cal{O}} (100~{\rm GeV})$ by our setting,  the magnitude of its
elements is at the order of $10^{-6}$.

Next we consider the sneutrino sector of the extension. One particular feature about the sneutrinos is that the lightest $\tilde{\nu}_R$-dominated sneutrino
can act as a viable DM candidate\footnote{Note that although the $\bar{\lambda}_{\nu} \hat{s} \hat{\nu} \hat{\nu}$ term in the superpotential violates lepton
number by $\Delta L = 2$, it does not spoil $R$-parity, which is defined by $R \equiv (-1)^{3 B + L + 2 S}$ with $B$, $L$ and $S$ denoting baryon number, lepton number and
spin of the involved field respectively, since $\Delta L$ is an even number. One can check this conclusion by studying the interactions induced by the term.},
and the advantage of such a choice over the customary neutralino DM is that, if $v_s$ is not excessively large (e.g. less than several {\rm TeV}),
the couplings of the sneutrino pair with SM particles are always weak due to its singlet nature. This causes its scattering with nucleon suppressed greatly, and thus alleviates the constraints of the DM DD experiments on the theory~\cite{Cao:2018iyk}. Throughout this work, we only discuss the case with a sneutrino DM since it corresponds to much broader parameter space allowed by current experiments.

For the neutrino/sneutrino sector, if one resorts the neutrino oscillations solely to the non-diagonality of the Yukawa coupling $Y_\nu$, $\bar{\lambda}_\nu$ is flavor diagonal.
If one further takes the soft breaking parameters $\bar{m}_{\tilde{l}}$ (slepton soft breaking mass), $\bar{m}_{\tilde\nu}$, $\bar{A}_{\lambda_{\nu}}$ and $\bar{A}_\nu$  to be flavor diagonal, the flavor mixings of the sneutrinos are extremely suppressed by the off-diagonal elements of $Y_\nu$, and it is a good approximation to only consider one generation sneutrinos in studying the properties of the sneutrino DM~\cite{Cao:2018iyk}.  In our discussion, we assume the sneutrino DM carrying a $\tau$ flavor, which is motivated by the fact that in some fundamental supersymmetric theories with supersymmetry broken at a high energy scale by certain mechanisms, the third generation sfermions are usually lighter than the other generation ones due to the renormalization group effects\footnote{From the perspective of collider phenomenology, the hypothesis predicts that charged supersymmetric particles decay ultimately into $\tau$ leptons in some popular cases. Detecting such a signal at the LHC is more difficult than the signal containing electrons or muon leptons, which can be learnt from the latest search for sleptons at the LHC (see~\cite{Aad:2019vnb} for $\tilde{e}_{L,R}$/$\tilde{\mu}_{L,R}$ and~\cite{Aad:2019byo} for $\tilde{\tau}_{L,R}$, as well as~\cite{Aad:2019qnd,Sirunyan:2019mlu} for compressed sparticle spectrum case). This makes the extension readily consistent with the results of the LHC in searching for supersymmetry. }. As a result, the parameters of the first two generation sneutrinos are irrelevant to our discussion. We use the symbols $m_{\tilde{l}}$, $\lambda_\nu$, $A_{\lambda_\nu}$ and $m_{\tilde{\nu}}$ to denote the 33 elements of the matrix $\bar{m}_{\tilde{l}}$, $\bar{\lambda}_\nu$, $\bar{A}_{\lambda_\nu}$ and $\bar{m}_{\tilde{\nu}}$ respectively and treat all these parameters as real numbers. Then after decomposing the sneutrino field into CP-even and CP-odd parts
\begin{equation}
\tilde{\nu}_L \equiv \frac{1}{\sqrt{2}}(\tilde{\nu}_{L1} + i
\tilde{\nu}_{L2}) ,
\quad\quad
\tilde{\nu}_R \equiv \frac{1}{\sqrt{2}}(\tilde{\nu}_{R1} + i \tilde{\nu}_{R2}) ,
\end{equation}
the sneutrino mass terms are written as~\cite{Cao:2018iyk}
\begin{eqnarray}
&&\frac{1}{2}(\tilde{\nu}_{Li}, \tilde{\nu}_{Ri})
\left(
\begin{array}{cc}
m_{L\bar{L}}^2           &  \pm m_{LR}^2+m_{L\bar{R}}^2 \\
\pm m_{LR}^2+m_{L\bar{R}}^2  &  m_{R\bar{R}}^2 \pm 2m_{RR}^2 \\
\end{array}
\right)
\left(
\begin{array}{c}
\tilde{\nu}_{Li}  \\
\tilde{\nu}_{Ri} \\
\end{array}
\right),   \label{Sneutrino-mass-matrix}
\end{eqnarray}
where $i=1, 2$ denote different CP states, the minus signs in the matrix are for the CP-odd states, and
\begin{eqnarray}
m_{L\bar{L}}^2
&\equiv& m_{\tilde{l}}^2 + |Y_\nu v_u|^2 + \frac{1}{8} (g_1^2 + g_2^2) (v_d^2 - v_u^2), \nonumber \\
m_{LR}^2
&\equiv& 2 Y_\nu v_u \left(\lambda v_s \right)^{\ast}, \nonumber\\
m_{L\bar{R}}^2
&\equiv& Y_\nu\left(-\lambda v_s v_d \right)^{\ast}
+ Y_\nu A_{Y_\nu} v_u , \nonumber\\
m_{R\bar{R}}^2
&\equiv& m_{\tilde{\nu}}^2 +|2\lambda_{\nu} v_s |^2
+ |Y_\nu v_u|^2 , \nonumber\\
m_{RR}^2
&\equiv& \lambda_\nu \left( A_{\lambda_\nu} v_s+(
\kappa v_s^2-\lambda v_d v_u )^{\ast} \right) .
\end{eqnarray}
Eq.(\ref{Sneutrino-mass-matrix}) indicates that the chiral mixings of the sneutrino fields are proportional to $Y_{\nu}$, and hence can be ignored safely. So the
sneutrino mass eigenstate coincides with the chiral state. It also indicates that, due to the presence of lepton number violating interactions in the superpotential,
the CP-even and -odd components of the right-handed sneutrino field are usually not degenerate in mass, and consequently the sneutrino DM has a definite CP number.
For more discussion about the property of the sneutrino DM, one can see our previous work~\cite{Cao:2018iyk}.

Possible annihilation channels of the sneutrino DM include~\cite{Cerdeno:2008ep,Cerdeno:2009dv}
\begin{itemize}
\item[(1)] $\tilde{\nu}_1 \tilde{H} \rightarrow X Y$ and $\tilde{H} \tilde{H}^\prime \rightarrow X^\prime Y^\prime$ with $\tilde{H}$ and $\tilde{H}^\prime$ denoting any Higgsino dominated neutralino or chargino and $X^{(\prime)}$ and $Y^{(\prime)}$ representing any lighter state.
	These annihilation channels are called coannihilation in literature~\cite{Coannihilation,Griest:1990kh}, and they are important only when
	the mass splitting between $\tilde{H}$ and $\tilde{\nu}_1$ is less than about $10\%$. As pointed out by the Bayesian analysis of the model in~\cite{Cao:2018iyk}, this channel is the most important annihilation mode.
\item[(2)] $\tilde{\nu}_1 \tilde{\nu}_1 \rightarrow s s^\ast $
	via the $s$-channel exchange of a Higgs boson, the $t/u$-channel exchange of a sneutrino, and any relevant scalar quartic couplings
     with $s$ denoting a light Higgs boson. This is the second important annihilation channel of the DM.
\item[(3)]  $\tilde{\nu}_1 \tilde{\nu}_1 \rightarrow V V^\ast$, $Vs$, $f \bar{f} $ with $V$ and $f$ denoting a vector boson ($W$ or $Z$) and
	a SM fermion, respectively.  This kind of annihilations proceeds via the $s$-channel exchange of a CP-even Higgs boson.
\item[(4)]  $\tilde{\nu}_1 \tilde{\nu}_1 \rightarrow \nu_R \bar{\nu}_R$ via the $s$-channel exchange of a Higgs boson and the $t/u$-channel exchange of a neutralino.
\item[(5)] $\tilde{\nu}_1 \tilde{\nu}_1^\prime \rightarrow A_i^{(\ast)} \rightarrow X Y$ and $\tilde{\nu}_1^\prime \tilde{\nu}_1^\prime \rightarrow X^\prime Y^\prime$ with $\tilde{\nu}_1^\prime$ denoting a sneutrino with an opposite CP number to that of $\tilde{\nu}_1$.
	These annihilation channels are important in determining the relic density only when the CP-even and -odd states are nearly
	degenerate in mass.
\end{itemize}
The expressions of $\sigma v$ for some channels are presented in \cite{Cerdeno:2009dv}. One can learn from them that the parameters in sneutrino sector, such as $\lambda_\nu$, $A_\nu$ and $m_{\tilde{\nu}}^2$, as well as the parameters in the Higgs sector,  are involved in the annihilations.

Note that the above introduction reveals the fact that the singlet Higgs field in the seesaw extension of the NMSSM plays an important roles in both Higgs physics and DM physics, e.g. besides being responsible for $\mu$ term and affecting Higgs mass spectrum, it also accounts for right-handed neutrino masses and sneutrino DM annihilation. This feature makes the theory quite distinct from its corresponding extension of the MSSM.

\subsection{Formula for the \texorpdfstring{$b\bar{b}$}{} and \texorpdfstring{$\gamma\gamma$}{} signals}

In the seesaw extension of the NMSSM, the singlet-dominated CP-even $h_1$ may account for both excesses. In order to illustrate this point, let's first
look at the analytic expression of the signal strengths for the excesses in the narrow width approximation. The diphoton signal strength normalized to its SM prediction is given by
\begin{eqnarray}
	\mu_{\rm CMS}|_{m_{h_1} \simeq 96~{\rm GeV}} &=&
  \frac{\sigma_{\rm SUSY}(p p \to h_1)}
       {\sigma_{\rm SM}(p p \to h_1 )} \times
       \frac{{\rm Br}_{\rm SUSY}(h_1 \to \gamma \gamma)}
       {{\rm Br}_{\rm SM}(h_1 \to \gamma \gamma)} \nonumber  \\
& \simeq &  \frac{\sigma_{\rm SUSY, ggF}(pp \to h_1)}{\sigma_{\rm SM, ggF}(pp \to h_1)} \times
       \frac{{\rm \Gamma}_{\rm SUSY}(h_1\to \gamma \gamma)}{\Gamma_{\rm SUSY}^{\rm tot}\times{\rm Br}_{\rm SM}(h_1 \to \gamma \gamma)}  \nonumber \\
& \simeq & \left|C_{h_1 g g}\right|^2 \times
  \frac{{\rm \Gamma}_{\rm SUSY}(h_1\to \gamma \gamma)}{\Gamma_{\rm SUSY}^{\rm tot}}
   \times \frac{1}{1.43 \times 10^{-3}},   \label{Diphoton-strength}
\end{eqnarray}
where the mass of the Higgs boson $h_1$ (denoted by $m_{h_1}$) is fixed around $96 {\rm GeV}$, and the subscript SUSY (SM) denotes the predictions of the Type-I NMSSM
(SM) on the inclusive production rate of $h_1$, its decay branching ratio into $\gamma \gamma$ and its width, which are labelled as $\sigma( p p \to h_1)$,
${\rm Br} (h_1 \to \gamma \gamma)$ and $\Gamma$ respectively. As shown in the experimental analysis~\cite{CMS-PAS-HIG-14-037}, the production rate
$\sigma( p p \to h_1)$ is mainly contributed
by gluon fusion (ggF) process, vector boson fusion (VBF) process, vector boson associated production (VH) as well as $t\bar{t} h_1$ production.
Among these contributions, the ggF process is the main one in the SM (which contributes about $86\%$ of the signal~\cite{CMS-PAS-HIG-14-037}) and also in the Type-I NMSSM (see footnote 4 below), so we approximate $\sigma( p p \to h_1)$ in the first equation by $\sigma_{\rm ggF} ( p p \to h_1)$ in the second step of the formula.
In the final expression, $C_{h_1 g g}$ represents the SUSY prediction of $h_1 g g$ coupling which is normalized to its SM prediction, and in the leading order approximation
it is equal to the ratio $\sigma_{\rm SUSY, ggF}(pp \to h_1)/\sigma_{\rm SM, ggF}(pp \to h_1)$\footnote{Note that since $C_{h_1 V V} \simeq C_{h_1 t \bar{t}} \simeq C_{h_1 g g}$ (see following discussion), $\sigma_{\rm SUSY, ggF}/\sigma_{\rm SM, ggF} \simeq \sigma_{\rm SUSY, VBF}/\sigma_{\rm SM, VBF} \simeq \sigma_{\rm SUSY, VH}/\sigma_{\rm SM, VH} \simeq \sigma_{\rm SUSY, t\bar{t}h_1}/\sigma_{\rm SM, t\bar{t}h_1} \simeq C_{h_1 g g}$. This implies that the ggF process is still the dominant one in contributing to the cross section $\sigma ( p p \to h_1)$ in the Type-I NMSSM, and $C_{h_1 g g}$ in the final expression of Eq.(\ref{Diphoton-strength}) can be treated as an approximation of the ratio $\sigma_{\rm SUSY}(pp \to h_1)/\sigma_{\rm SM}(pp \to h_1)$. The goodness of this approximation is not sensitive to the fraction of the ggF contribution to the total signal. }.  $\Gamma_{\rm SUSY}^{\rm tot} = \Gamma_{\rm SUSY}(h_1 \to b\bar{b}) + \Gamma_{\rm SUSY}(h_1\to c\bar{c}) + \cdots$ denotes the SUSY prediction on the total width of $h_1$, and $1.43\times10^{-3}$ corresponds to the branching ratio of $h_1 \to \gamma \gamma$ in the SM for $m_{h_1} = 96 {\rm GeV}$, which includes all known higher-order QCD corrections and is calculated by LHC Higgs Cross Section Working Group~\cite{Heinemeyer:2013tqa}. In getting the value of $\mu_{\rm CMS}$ by the final expression, we include all one-loop contributions (which are induced by quarks and squarks) to $C_{h_1 g g}$ and all leading order contributions to $\Gamma_{\rm SUSY}(h_1\to \gamma \gamma)$ and $\Gamma_{\rm SUSY}^{\rm tot}$.
The signal strength of the $b\bar{b}$ excess, $\mu_{\rm LEP}$, is defined in a similar way to $\mu_{\rm CMS}$, and is given by
\begin{eqnarray}
	\mu_{\rm LEP}|_{m_{h_1} \simeq 96~{\rm GeV}} &=&
  \frac{\sigma_{\rm SUSY}(e^+e^-\to Z h_1)}
       {\sigma_{\rm SM}(e^+e^-\to Z h_1)} \times
       \frac{{\rm Br}_{\rm SUSY}(h_1\to b\bar{b})}
       {{\rm Br}_{\rm SM}(h_1 \to b\bar{b})} \nonumber \\
  & = &
  \left|C_{h_1 V V}\right|^2 \times
  \frac{{\rm \Gamma}_{\rm SUSY}(h_1\to b\bar{b})}{\Gamma_{\rm SUSY}^{\rm tot}}
   \times \frac{1}{0.799}   \label{muLEP}
\end{eqnarray}
where $C_{h_1 VV}$ is the normalized coupling of $h_1$ with vector bosons, and $0.799$ is value of ${\rm Br}(h_1 \to b\bar{b})$ in the SM presented by the Higgs Cross Section Working Group~\cite{Heinemeyer:2013tqa}.

From the formulas of $\mu_{\rm CMS}$ and $\mu_{\rm LEP}$, one can learn two facts. One is that both strengths are expressed in term of the ratio $\sigma_{\rm SUSY}/\sigma_{\rm SM}$, and consequently the QCD correction to the numerator and the denominator will cancel. This is beneficial to reduce the theoretical uncertainty in predicting
the strengths. The other is that, in order to explain both excesses by a singlet Higgs boson, $h_1$ can not be CP-odd because a CP-odd Higgs boson does not couple with $ZZ$ and consequently it has no contribution to the $b \bar{b}$ excess. If alternatively one just want to explain the diphoton excess, the Higgs boson can be either CP-even or CP-odd.

Next we scrutinize the involved couplings. Since current LHC data have required the properties of the discovered boson to highly mimic
those of the SM Higgs boson and meanwhile colored sparticles heavier than about $1{\rm TeV}$, we have following approximation for the normalized couplings of $h_1$~\cite{Cao:2016uwt}
\begin{eqnarray}
C_{h_1 t \bar{t}} &\simeq& - V_{11} \cot \beta + V_{12}, \quad C_{h_1 g g} \simeq C_{h_1 t \bar{t}}, \\ \nonumber
C_{h_1 b \bar{b}} &\simeq&  V_{11} \tan \beta + V_{12}, \quad C_{h_1 V V} = V_{12},   \label{Couplings-approximation}
\end{eqnarray}
where $V_{ij}$ with $i,j=1,2,3$ denotes the element of the rotation matrix to diagonalize the mass matrix in Eq.(\ref{CP-even Mass}). As for $C_{h_1 \gamma \gamma}$, besides the top quark- and W-mediated loops, it is also contributed by chargino loops and charged Higgs loop, i.e. $C_{h_1 \gamma \gamma} = C_{h_1 \gamma \gamma}^t + C_{h_1 \gamma \gamma}^{W^\pm} + C_{h_1 \gamma \gamma}^{\tilde{\chi}^\pm} + C_{h_1 \gamma \gamma}^{H^\pm}$. Although the charged Higgs loop is usually negligible since it is mediated by a heavy scalar particle~\cite{King:2012tr}, the Higgsino-dominated chargino loop may play a role in enhancing $\Gamma (h_1 \to \gamma \gamma)$, which can be inferred from~\cite{Choi:2012he}
\begin{eqnarray}
C_{h_1 \gamma \gamma}^{\tilde{\chi}^\pm} &\simeq & \left ( \frac{2}{9} A_{1/2} (\tau_t) - \frac{7}{8} A_1 (\tau_W) \right )^{-1} \times \frac{\lambda v}{6 |\mu|} \left ( 1 + \frac{7}{30} \frac{m_{h_1}^2}{4 \mu^2} \right ) V_{13} \nonumber \\
& \simeq & -1.37 \times \frac{\lambda v}{6 |\mu|} \left ( 1 + \frac{7}{30} \frac{m_{h_1}^2}{4 \mu^2} \right ) V_{13} \quad \quad \quad {\rm for}~~ m_{h_1} = 96~{\rm GeV},
\end{eqnarray}
with $A_{1/2}$ and $A_1$ being loop functions with $\tau_i = m_{h_1}^2/(4 m_i^2)$. For example, if $\lambda > \frac{1}{5.6 |V_{13}|} \frac{|\mu|}{100~{\rm GeV}}$, one has $|C_{h_1 \gamma \gamma}^{\tilde{\chi}^\pm}| \gtrsim 0.1$. This is not a negligible number since $C_{h_1 \gamma \gamma} \simeq 0.3$ can account for the diphoton excess at $1 \sigma$ level (see the results in Fig.\ref{fig2}).

From these formulas, one can learn following facts
\begin{itemize}
\item If the theory is used to explain the excesses, the preferred mass spectrum is $m_{h_1} \simeq 96~{\rm GeV}$,  $m_{h_2} \simeq 125~{\rm GeV}$ and $m_{h_3} \simeq m_{H^\pm} \gtrsim 500~{\rm GeV}$. Since the splitting between $m_{h_1}$ and $m_{h_3}$ is much larger than that between $m_{h_1}$ and $m_{h_2}$, $V_{12} \gg V_{11}$ is valid for most cases. So one can conclude that $C_{h_1 V V} \simeq C_{h_1 t \bar{t}} \simeq C_{h_1 g g} \simeq V_{12}$ and $C_{h_1 \gamma \gamma}^t + C_{h_1 \gamma \gamma}^W \simeq V_{12}$. This estimation is helpful to understand the strengths.
\item $C_{h_1 b \bar{b}}$ may be significantly smaller than $C_{h_1 t \bar{t}}$ due to the cancellation between $V_{11} \tan \beta$ and $V_{12}$. In this case, $\Gamma_{\rm SUSY}^{\rm tot}$ is reduced greatly, but it does not change the fact that $h_1 \to b\bar{b}$ is the dominant decay channel of $h_1$ since the Yukawa coupling of $h_1$ with bottom quark is usually much larger than its couplings with the other light quarks and leptons.
\item An uncertainty of $10\%$ in $C_{h_2 V V}$ measurement by the latest Higgs data at the LHC~\cite{ATLAS:2019slw} implies that $|C_{h_1 V V}|^2 \lesssim 0.2$. This size is large enough to produce the central value of the $b \bar{b}$ excess because ${\rm Br}(h_1 \to b \bar{b})$ is insensitive to $C_{h_1 b \bar{b}}$ unless it is suppressed too much (see Eq.\ref{muLEP}).
\item A moderately large $C_{h_1 \gamma \gamma}^{\tilde{\chi}^\pm}$ (compared with the top- and W-loop contribution) together with a suppressed $C_{h_1 b \bar{b}}$  (relative to $C_{h_1 t \bar{t}}$) are favored to explain the diphoton excess\footnote{Note that similar conditions to enhance the ratio ${\rm Br} (h_1 \to \gamma \gamma)/{\rm Br}_{\rm SM} (h_1 \to \gamma \gamma)$ in supersymmetric theories have been obtained in~\cite{Ellwanger:2010nf,Cao:2011pg}.}. This can be understood by the fact
    \begin{eqnarray}
    \frac{\mu_{\rm CMS}}{\mu_{\rm LEP}} = \frac{C_{h_1 g g}^2}{C_{h_1 V V}^2}  \times  \frac{C_{h_1 \gamma \gamma}^2}{C_{h_1 b \bar{b}}^2} \simeq  \frac{C_{h_1 \gamma \gamma}^2}{C_{h_1 b \bar{b}}^2} \simeq \left ( \frac{V_{12} + C_{h_1 \gamma \gamma}^{\tilde{\chi}^\pm}}{V_{11} \tan \beta + V_{12}} \right )^2 \sim 5,
    \end{eqnarray}
     where the number 5 is obtained from the central values of the excesses in Eq.(\ref{mu-Excesses}). This formula reveals that if $C_{h_1 \gamma \gamma}^{\tilde{\chi}^\pm} \simeq 0$, the condition $V_{11} \tan \beta \simeq -0.55  V_{12}$ must be satisfied to predict the excesses. This will put strong constraint on the parameter space of the Type-I NMSSM, while a varying $C_{h_1 \gamma \gamma}^{\tilde{\chi}^\pm}$ can relax the correlation.
\end{itemize}

\section{Explanations of the excesses}  \label{Section-excess}

In this section we attempt to explain the excesses in the NMSSM with the Type-I seesaw mechanism.
We utilize the package \textsf{SARAH-4.11.0}~\cite{sarah-1,sarah-2,sarah-3} to build the model, the codes  \textsf{SPheno-4.0.3}~\cite{spheno} and \textsf{FlavorKit}~\cite{Porod:2014xia} to generate particle spectrum and compute low energy
flavor observables respectively, and the package \textsf{HiggsBounds-5.3.2}~\cite{HiggsBounds} to implement the constraints
from the direct search for extra Higgs bosons at LEP, Tevatron and LHC. For some benchmark settings, we also use
the package \textsf{MicrOMEGAs 4.3.4}~\cite{Belanger:2013oya,micrOMEGAs-1,micrOMEGAs-3}
to compute DM observables by assuming the lightest sneutrino as the only DM candidate in the universe.
In calculating the radiative correction to the Higgs mass spectrum, the code \textsf{SPheno-4.0.3}
only includes full one- and two-loop effects using a diagrammatic approach with vanishing external
momenta~\cite{spheno}. This leaves an uncertainty of about $2~{\rm GeV}$ for the SM-like Higgs boson mass.

\subsection{Strategy in scanning the parameter space}

Previous discussions indicate that only the parameters in the Higgs sector determine the $b\bar{b}$ and $\gamma \gamma$ signals.
We perform a sophisticated scan over these inputs and the soft trilinear coefficient $A_{t}$ for top squark (since this parameter can affect significantly
the Higgs mass spectrum by radiative corrections)  in following ranges\footnote{Note that we are not intend to perform a complete fit of the model to the excesses in this work, so we only select by experience part of its parameter space for study.}
\begin{equation}
\begin{split}\label{eq:Higgs-range}
&  0 < \lambda \leq 0.75,\quad 0 < \kappa \leq 0.75,\quad 1 \leq \tan{\beta} \leq 20,	\quad 100~{\rm GeV} \leq \mu \leq 600~{\rm GeV} ,\\
& 300~{\rm GeV}\leq A_\lambda \leq 2{\rm TeV},\quad -1{\rm TeV} \leq A_\kappa \leq 0 {\rm TeV},\quad |A_t| \leq 5{\rm TeV},
\end{split}
\end{equation}
where all the parameters are defined at the scale $Q=1{\rm TeV}$.  The other unimportant parameters are set as follows: $\lambda_\nu = 0.1$,
$M_1 = M_2=2{\rm TeV}$ and $M_3=5{\rm TeV}$ for gaugino soft breaking masses, and all soft breaking parameters in squark and slepton sectors
except $A_t$ are fixed at $2{\rm TeV}$, which are consistent with the results of the LHC search for sparticles.  In the scan,
we adopt the MultiNest algorithm in~\cite{Feroz:2008xx} with the flat distribution for the inputs
and {\it{nlive}} $= 10000$, and construct the likelihood function
	\begin{eqnarray}
	\mathcal{L}=\mathcal{L}_{\rm{excess}} \times  \mathcal{L}_{h_2, \rm{mass}} \equiv Exp\left [ -\frac{1}{2} \chi^{2}_{\rm H} \right ], \label{Excess-Likelihood}
	\end{eqnarray}
to guide the scan, where $\chi^2_{\rm H}=\chi^2_{\rm excess} + \chi^2_{h_2,{\rm mass}}$ with $\chi^2_{{\rm excess}}$ and $\chi^2_{h_2, \rm{mass}}$ denoting the $\chi^2$ function of the excesses and $m_{h_2}$ respectively and their forms given by\footnote{We assume relatively small total (theoretical and experimental) uncertainties for $m_{h_1}$ and $m_{h_2}$ in the study, i.e. $\Delta m_{h_1} = 0.2~{\rm GeV}$ and $\Delta m_{h_2} = 2~{\rm GeV}$, to ensure that the samples obtained in the scan focus on the case $m_{h_1} \simeq 96~{\rm GeV}$ and $m_{h_2} \simeq 125~{\rm GeV}$. Moreover, we do not include the coupling information of the discovered Higgs boson in the $\mathcal{L}$ because we want to get the best explanations to the excess instead of to perform a global fit of the model with all experimental data. This is vital in our calculation since so far the excesses are not very significant.}
\begin{eqnarray}
\chi^2_{\rm{excess}} &=& \left(\frac{m_{h_1}-96.0}{0.2}\right)^2 + \left( \frac{\mu_{\rm LEP}-0.117}{0.057} \right)^2 + \left( \frac{\mu_{\rm CMS}-0.6}{0.2} \right)^2, \\
\chi^2_{h_2, \rm{mass}} &=& \left(\frac{m_{h_2}-125.1}{2.0}\right)^2.
\end{eqnarray}
Note that the setting {\it{nlive}} in the MultiNest method denotes the number of active or live points used to determine the iso-likelihood contour in each iteration~\cite{Feroz:2008xx,Feroz:2013hea}. The larger it is, the more meticulous the scan becomes in surveying the parameter space.

In the scan, we also calculate following $\chi^2$ functions for each sample
\begin{itemize}
\item $\chi^2_{h_2, {\rm couplings}}$ for seven couplings of the discovered Higgs boson  in the $\kappa$-framework, which were recently obtained by ATLAS collaboration with $80~{\rm fb}^{-1}$ data. We assume no exotic decay of $h_2$, and use the coupling information for the scenario (a) in Table 11 of~\cite{ATLAS:2019slw} and its corresponding correlation matrix in Figure 38 of the same experimental report to calculate the $\chi^2_{h_2, {\rm couplings}}$. We do not include the theoretical uncertainty in calculating the couplings since they are much smaller than corresponding experimental uncertainty.
\item $\chi^2_B$ for the measurement of ${\rm Br} (B\to X_s \gamma)$ and ${\rm Br} (B_s \to \mu^+\mu^-)$, which takes the form~\cite{Tanabashi:2018oca}
\begin{equation*}
\chi^2_B = \frac{(B_\gamma - 3.43)^2}{0.4^2} +  \frac{(B_{\mu^+ \mu^-} - 3.11)^2}{1.2^2}
\end{equation*}
with $B_\gamma$ and $B_{\mu^+ \mu^-}$ denoting the theoretical prediction of ${\rm Br} (B \to X_s \gamma)$ and ${\rm Br} (B_s \to \mu^+\mu^-)$ in unit of $10^{-4}$ and $10^{-9}$, respectively.
\item $\chi^2_{\rm EW}$ for precision electroweak measurements $\epsilon_i$ ($i=1,2,3$)~\cite{Altarelli:1990zd,Altarelli:1991fk,Altarelli:1994iz} or equivalently $S$, $T$ and $U$ parameters~\cite{Peskin:1990zt,Peskin:1991sw}. We use the formulas for the self-energies of the gauge bosons $\gamma$, $W^\pm$ and $Z$ in~\cite{Cao:2008rc} to calculate these observables, and the fit results in~\cite{deBlas:2016ojx} to get the $\chi^2_{\rm EW}$.
\end{itemize}

\begin{figure*}[t]
		\centering
		\resizebox{1.\textwidth}{!}{
        \includegraphics[width=0.90\textwidth]{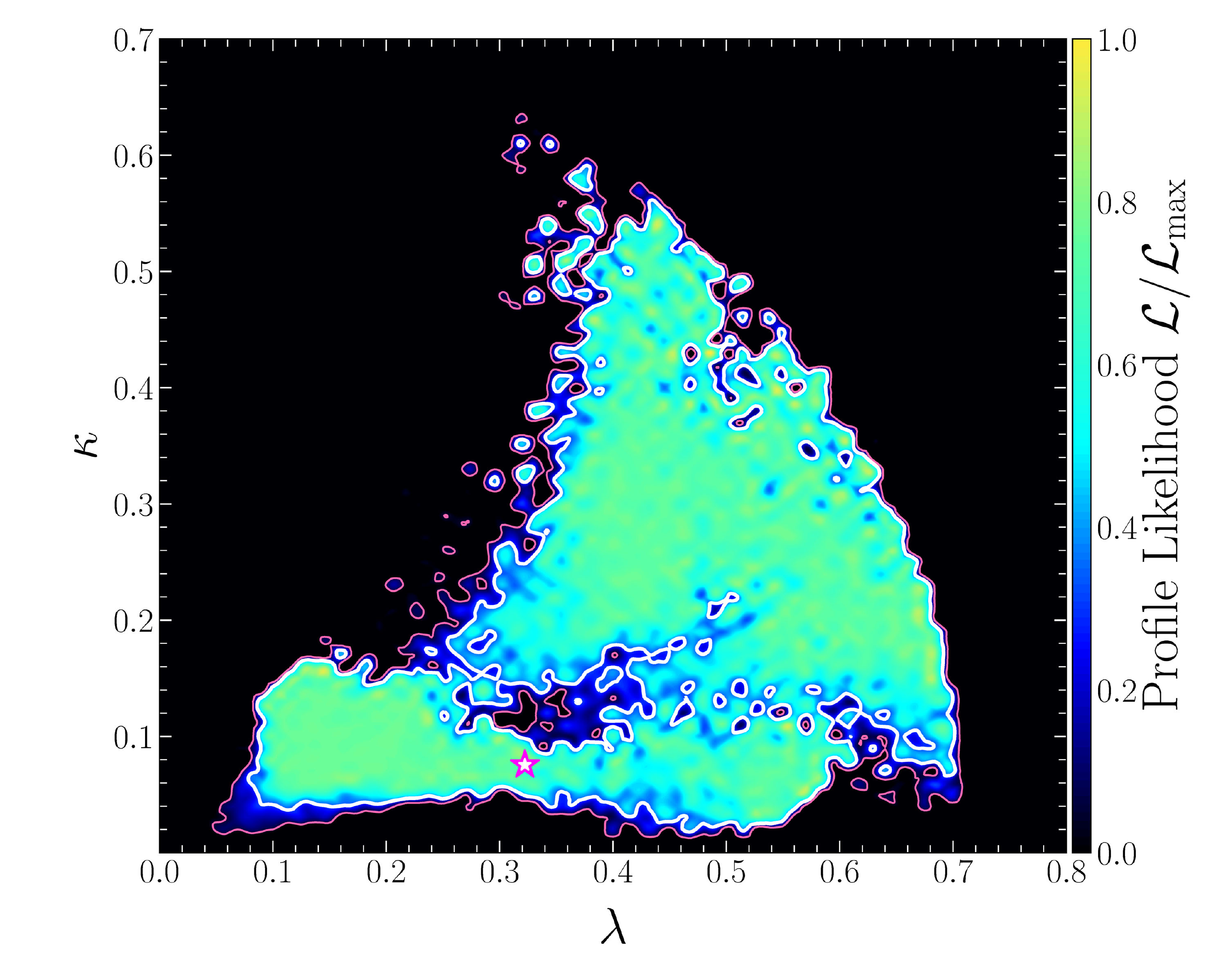}
        \includegraphics[width=0.90\textwidth]{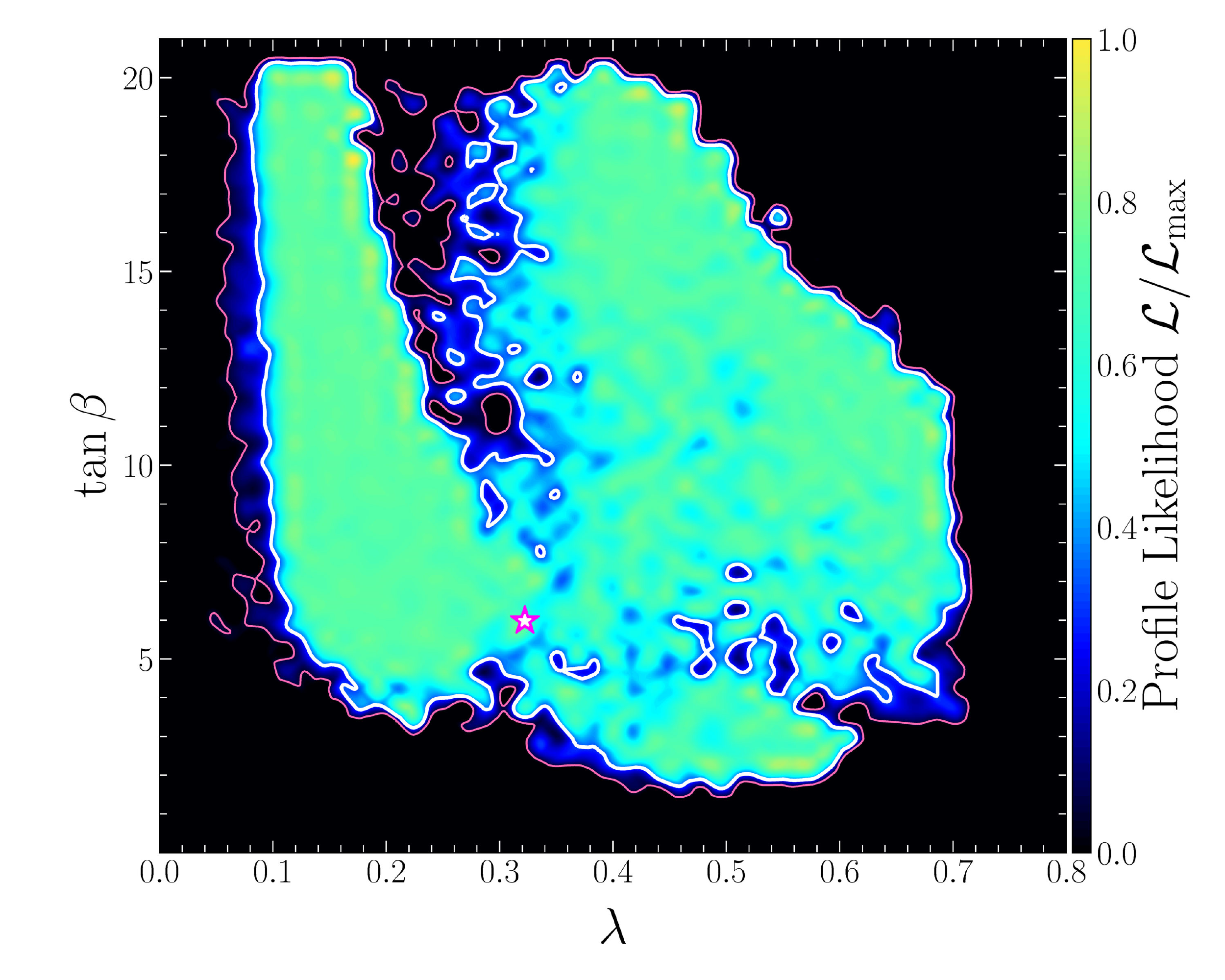}
        }

        \vspace{0.1cm}

		\resizebox{1.\textwidth}{!}{
        \includegraphics[width=0.90\textwidth]{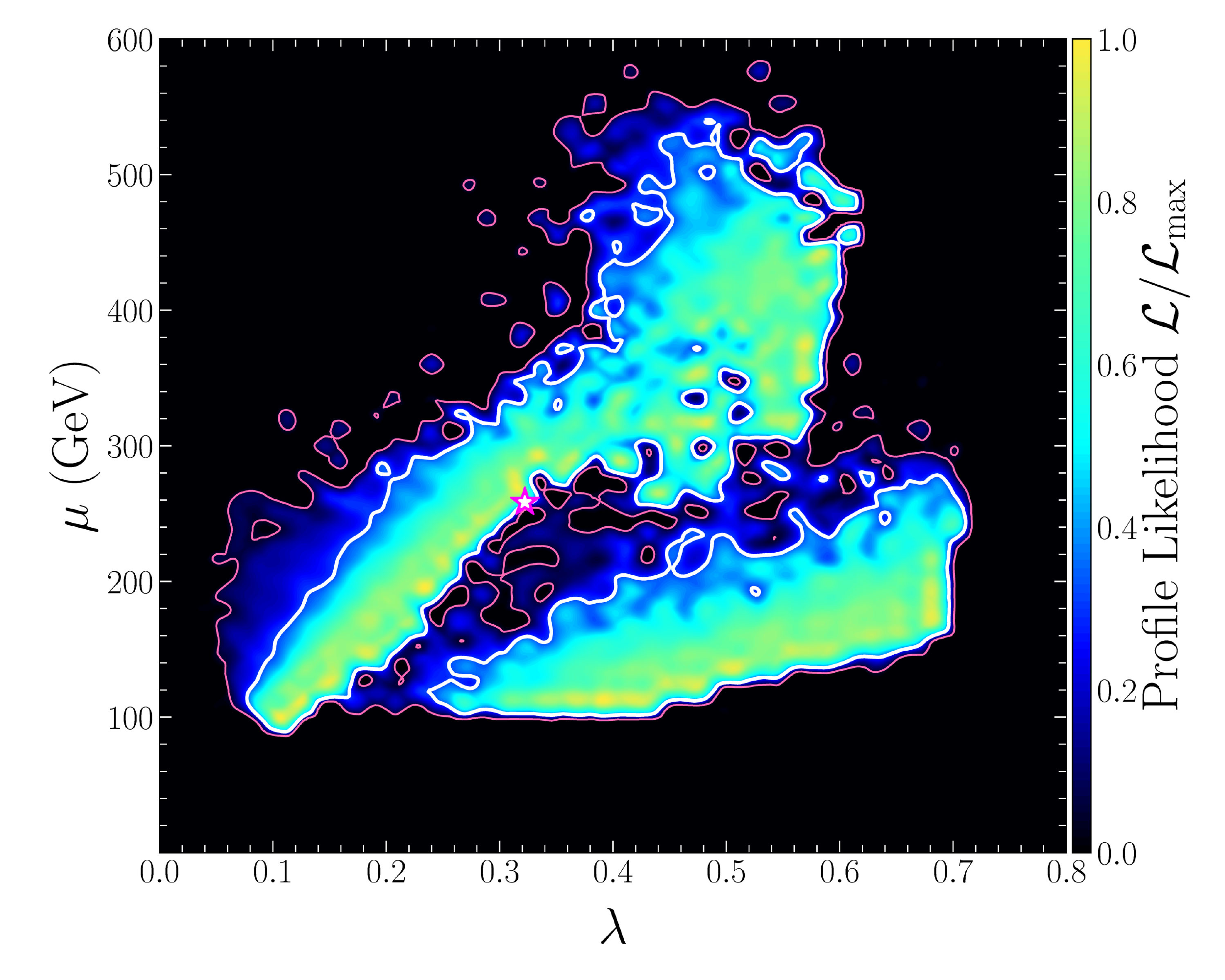}
        \includegraphics[width=0.90\textwidth]{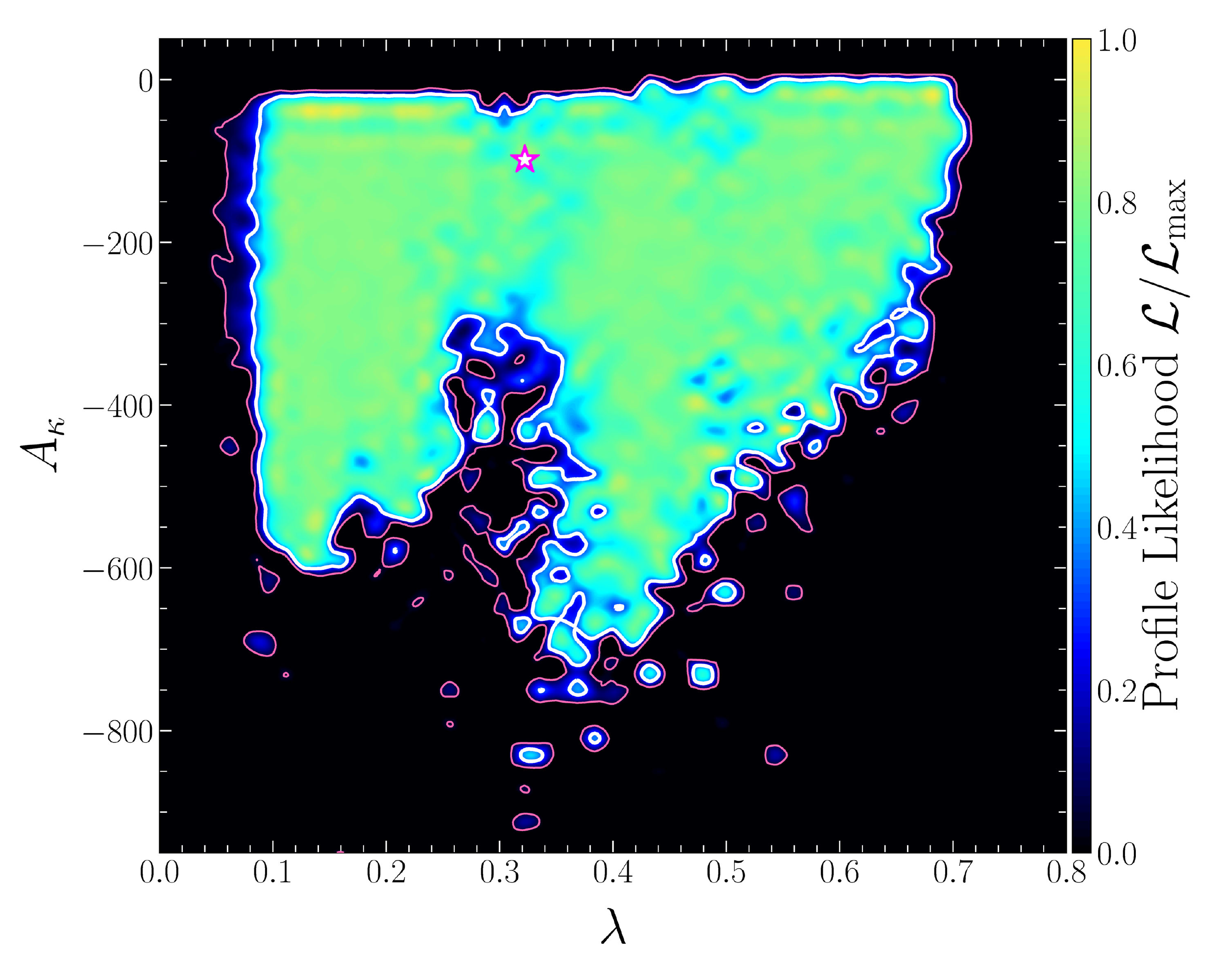}
        }
       \vspace{-0.3cm}
        \caption{Two dimensional profile likelihood of $\mathcal{L}$ in Eq.(\ref{Excess-Likelihood}), which are projected on $\kappa-\lambda$, $\tan \beta -\lambda$, $\mu-\lambda$ and $A_\kappa-\lambda$ planes respectively. Since $\chi^2_{\rm H, min} \simeq 0$ for the best point (marked by star symbol in the figure),
        the $1 \sigma$ boundary (white solid line) and the $2\sigma$ boundary (red line) correspond to $\chi^2_{\rm H} \simeq 2.3$ and $\chi^2_{\rm H} \simeq 6.18$, respectively. This figure reflects the preference of the excesses on the parameter space of the extended NMSSM. Note that all samples in this figure satisfy the conditions above Eq.(\ref{chi2-tot}), especially $\chi^2_{h_2, {\rm coupling}} \leq 14.1$, so they are consistent with the data for the discovered Higgs boson at $2\sigma$ level. \label{fig1} }
\end{figure*}	

\begin{figure*}[t]
		\centering
		\resizebox{1.\textwidth}{!}{
        \includegraphics[width=1.0\textwidth]{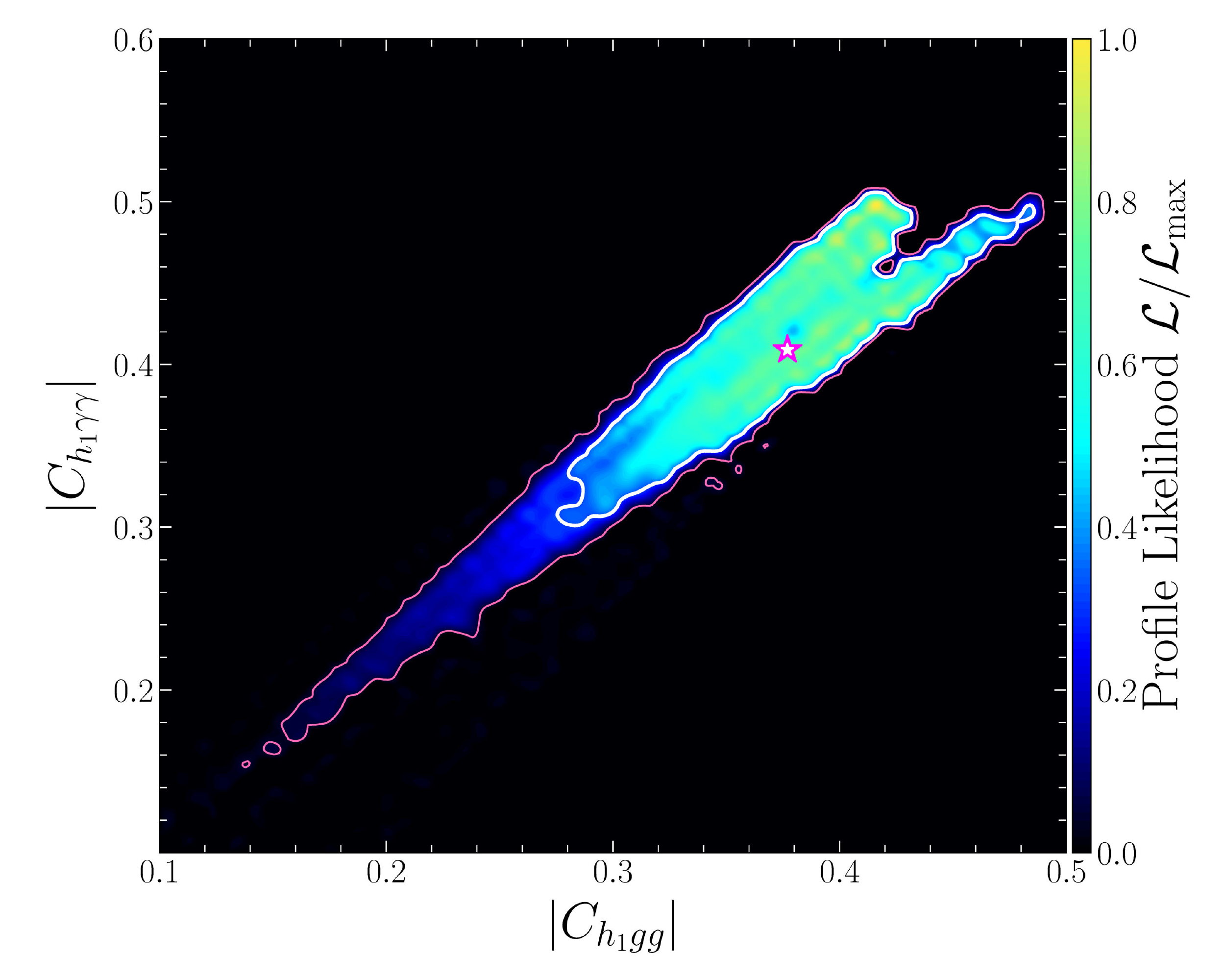}
        \includegraphics[width=1.0\textwidth]{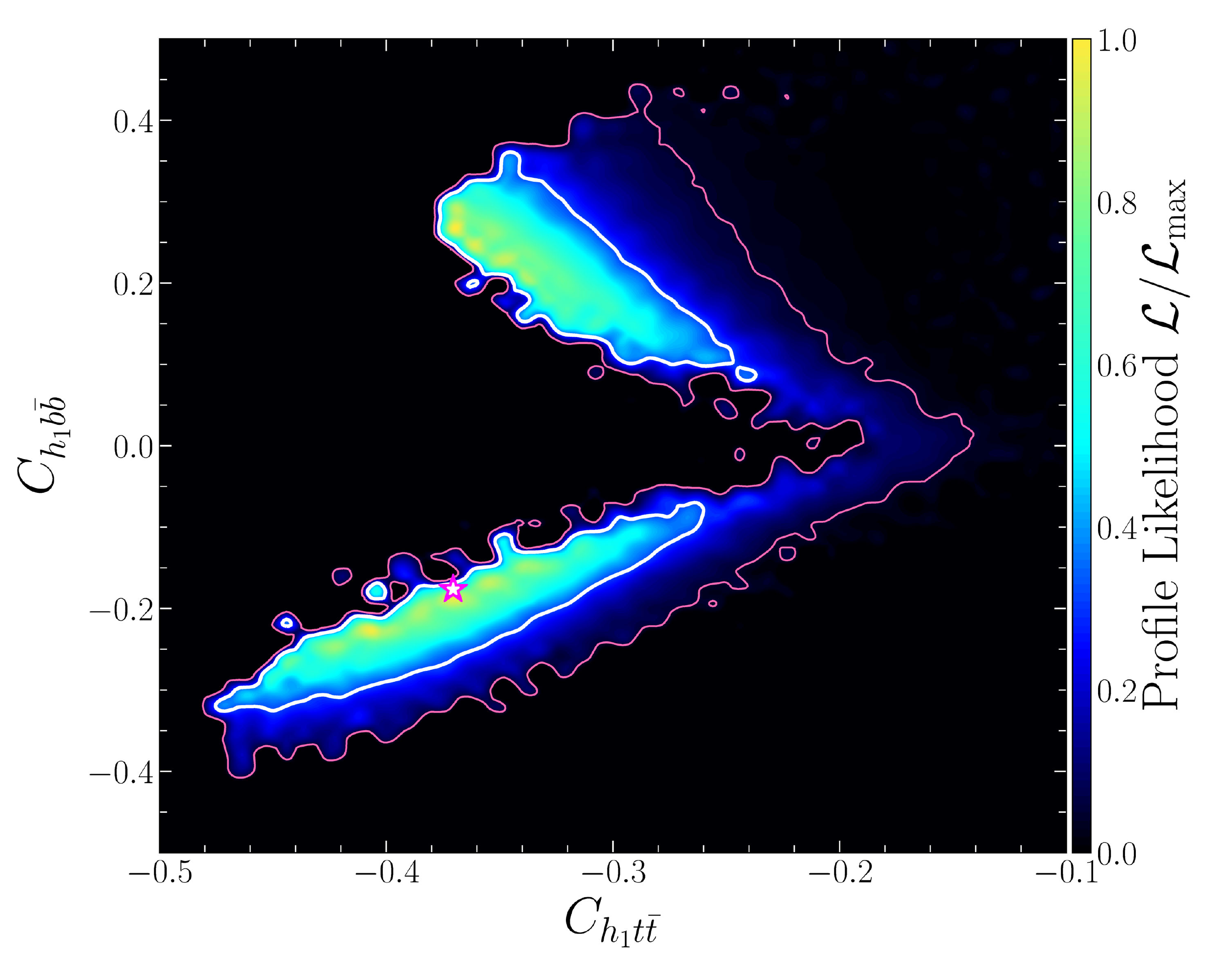}
        }

        \vspace{0.1cm}

		\resizebox{1.\textwidth}{!}{
        \includegraphics[width=1.0\textwidth]{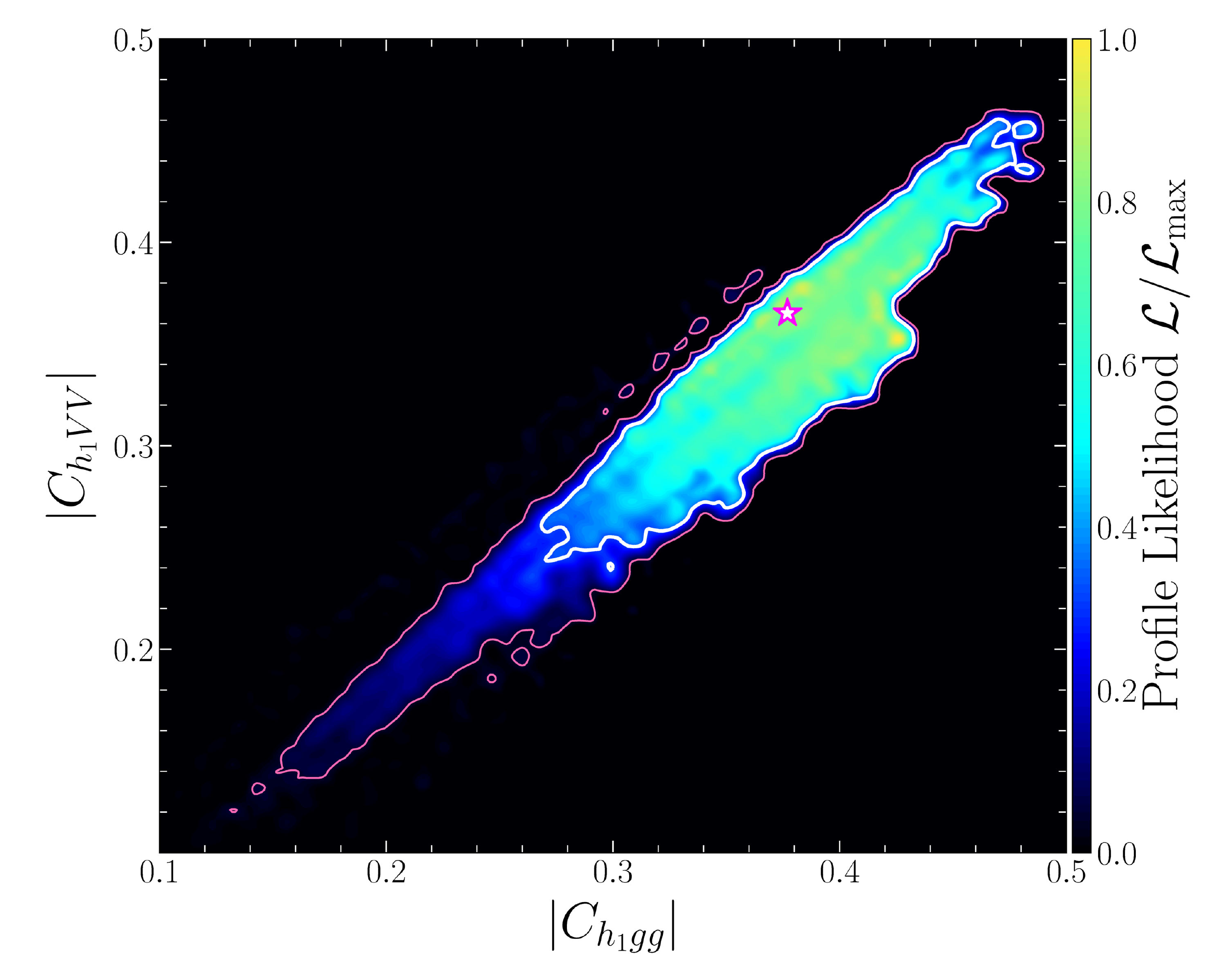}
        \includegraphics[width=1.0\textwidth]{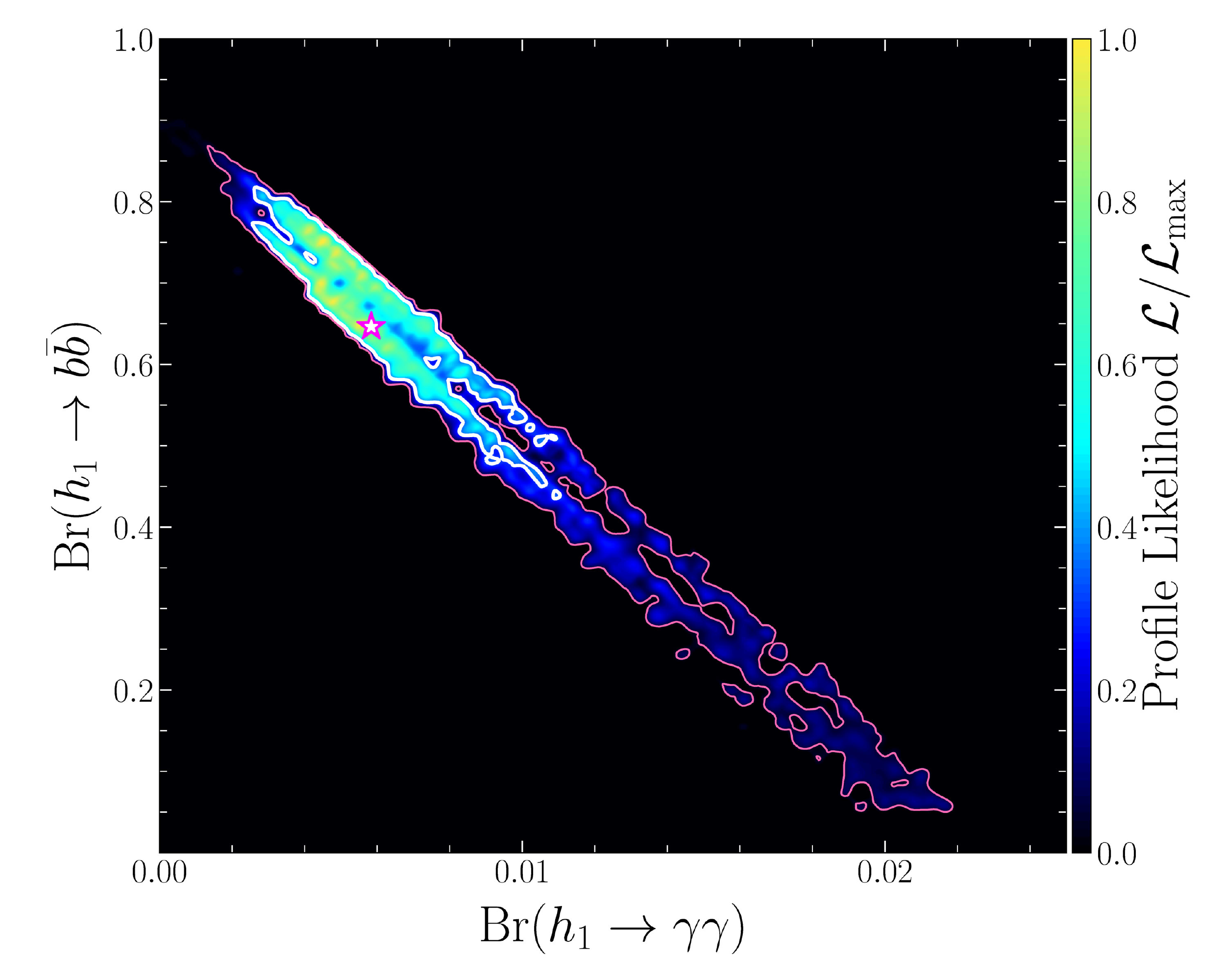}
        }
       \vspace{-0.3cm}
        \caption{Similar to Fig.\ref{fig1}, but projected on $|C_{h_1\gamma\gamma}|-|C_{h_1gg}|$, $C_{h_1b\bar{b}}-C_{h_1t\bar{t}}$,
        $|C_{h_1VV}|-|C_{h_1gg}|$ and ${\rm Br}(h_1\to\gamma\gamma)-{\rm Br}(h_1\to b\bar{b})$ planes, respectively.  \label{fig2} }
\end{figure*}	
\begin{figure*}[t]
		\centering
		\resizebox{1.\textwidth}{!}{
        \includegraphics[width=0.9\textwidth]{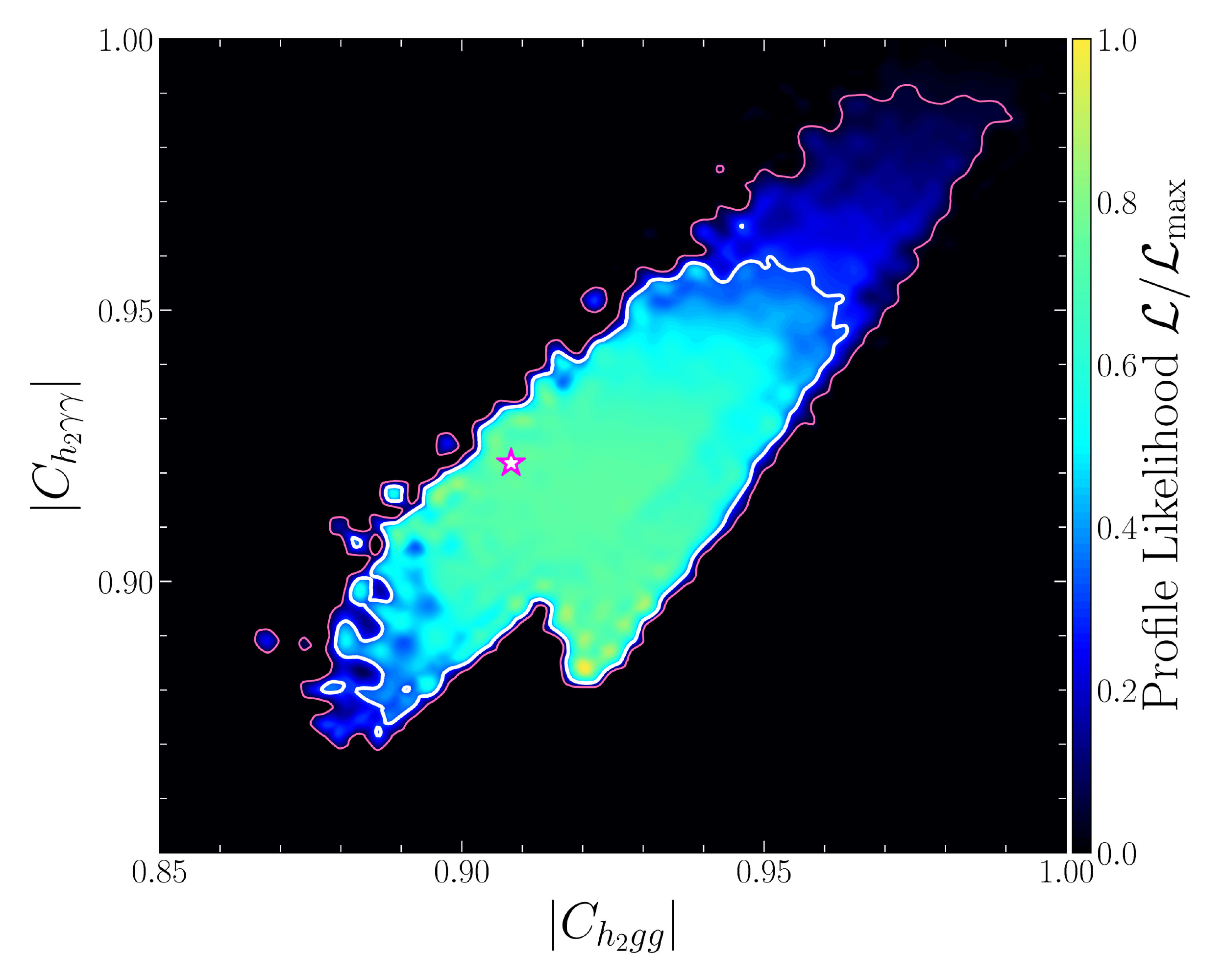}
        \includegraphics[width=0.9\textwidth]{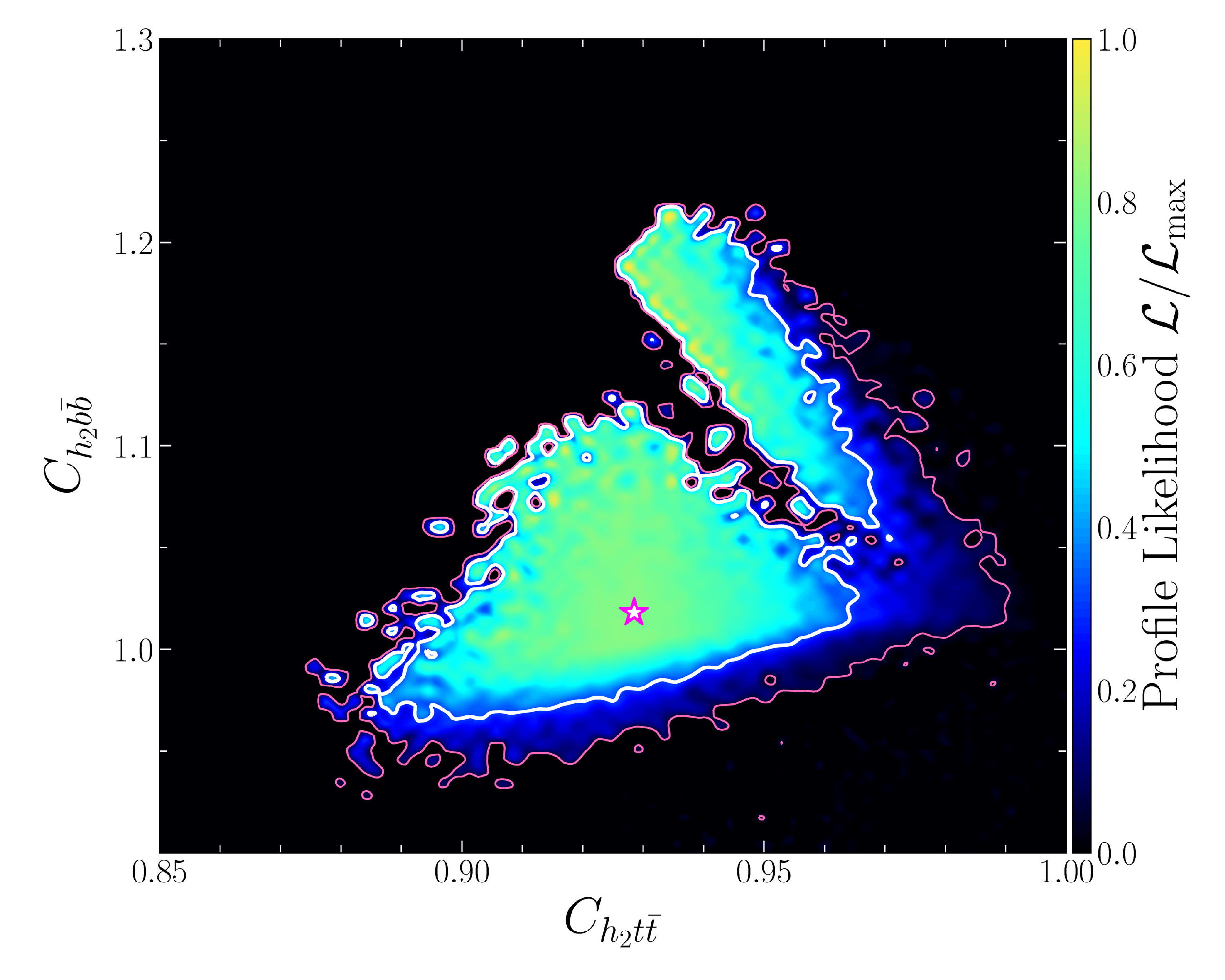}
        }
		\resizebox{1.\textwidth}{!}{
        \includegraphics[width=0.9\textwidth]{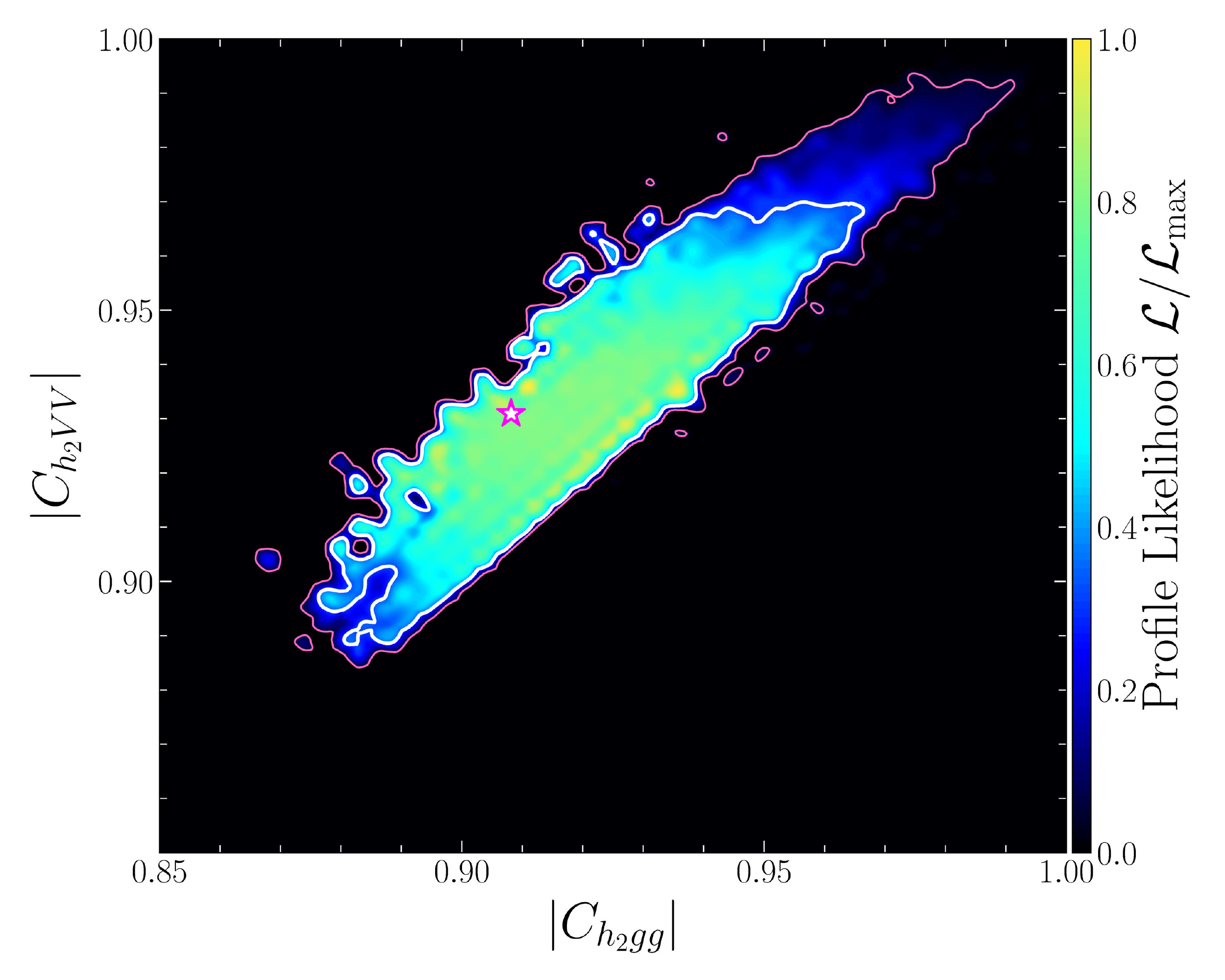}
        \includegraphics[width=0.9\textwidth]{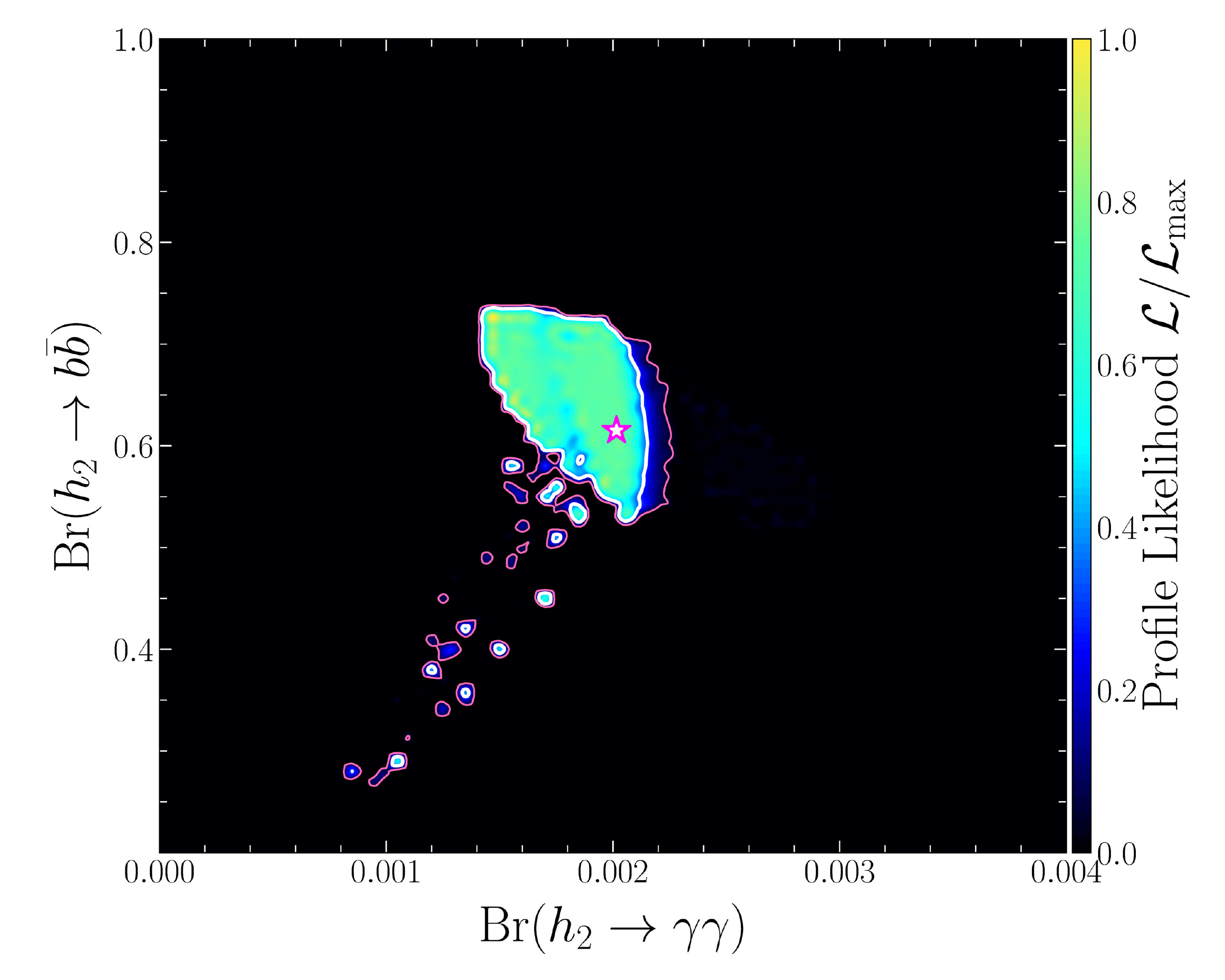}
        }
       \vspace{-0.4cm}
        \caption{Similar to Fig.\ref{fig1}, but projected on $C_{h_2 \gamma\gamma}-C_{h_2 gg} $, $C_{h_2 b\bar{b}}-C_{h_2 t\bar{t}}$,
        $C_{h_2 VV}- C_{h_2 gg} $ and ${\rm Br}(h_2 \to\gamma\gamma)-{\rm Br}(h_2 \to b\bar{b})$ planes, respectively.   \label{fig3} }
\end{figure*}	

In getting the explanations of the excesses, we refine the samples obtained in the scan by following conditions:
$m_{A_1}> m_{h_2}/2$ so that the discovered Higgs boson has no exotic decay,  $\lambda^2 + \kappa^2 \leq 0.5$ so that the theory keeps perturbative up to $10^{16}~{\rm GeV}$ scale~\cite{Miller:2003ay}, $\chi^2_{h_2, {\rm coupling}} \leq 14.1$ which is $95\%$ confidence level exclusion limit of the Higgs couplings for seven degrees of freedom and $\chi^2_{\rm tot} \leq 18.6$ with $\chi^2_{\rm tot}$ defined by\footnote{$\chi^2_{\rm tot}$ denotes a measure of the agreement between the theory and the total experimental data considered in this work. In this hypothesis, the goodness-of-fit measure $\chi^2_{\rm tot}$ obeys a $\chi^2$ distribution with $N_{\rm obs}-N_{\rm para}+1$ degree of freedom (d.o.f.). In our study, the d.o.f. is $16-7+1=10$, and $\chi^2_{\rm tot}=18.6$ corresponds to the upper limit of $\chi^2_{\rm tot}$ at $2\sigma$ confidence level.}
\begin{eqnarray}
\chi^2_{\rm tot} = \chi^2_{\rm{excess}} + \chi^2_{h_2, \rm{mass}} + \chi^2_{h_2, {\rm couplings}} + \chi^2_B + \chi^2_{\rm EW}.  \label{chi2-tot}
\end{eqnarray}
We also require the samples to survive the constraints from the \textsf{HiggsBounds}.

At this stage, we remind that, if one does not consider the constraints from DM physics and the relevant sparticle searches at the LHC, the Higgs physics of the NMSSM is same as that of the extended model. So one may also use the package \textsf{NMSSMTools}~\cite{Ellwanger:2004xm,Ellwanger:2005dv} to perform the scan. We compare the \textsf{NMSSMTools} with our toolkit, and find that their explanations of the excesses shown in following figures are similar, although the  \textsf{NMSSMTools} is somewhat faster than our toolkit in calculation.

\subsection{Numerical Results}

Based on the samples obtained in the scan, we plot the profile likelihoods (PL) of the $\mathcal{L}$ in Eq.(\ref{Excess-Likelihood}) on different planes\footnote{The frequentist PL is defined as the largest likelihood value in a certain parameter space~\cite{Fowlie:2016hew}. Given a likelihood function $\mathcal{L}$ defined in N-dimensional space $\Theta = (\Theta_1, \Theta_2, \cdots, \Theta_N)$, its two dimensional PL can be obtained by the procedure
\begin{eqnarray}
\mathcal{L}(\Theta_i,\Theta_j)=\mathop{\max}_{\Theta_1,\cdots,\Theta_{i-1},\Theta_{i+1},\cdots, \Theta_{j-1}, \Theta_{j+1},\cdots, \Theta_N}\mathcal{L}(\Theta). \nonumber
\end{eqnarray}
Obviously, the PL reflects the preference of a theory on the parameter space, and for a given point on $\Theta_i-\Theta_j$ plane, the value of $\mathcal{L}(\Theta_i,\Theta_j)$ represents the capability of the point in the theory to account for experimental data by varying the other parameters.}, where the color bar in Fig.\ref{fig1}, \ref{fig2} and \ref{fig3} represents the PL value relative to the best point marked by star symbol, and the white and pink solid lines are boundaries for $1\sigma$ and $2\sigma$ confidence intervals (CI), respectively. Fig.\ref{fig1} indicates that there are broad parameter space to explain the excesses and the large deviation between the $1\sigma$ and $2 \sigma$ boundaries on $\mu -\lambda$ plane reflects that the explanation is sensitive to the parameters $\lambda$ and $\mu$. Fig.\ref{fig2} shows that the magnitude of the normalized couplings of $h_1$ may reach 0.5 except $|C_{h_1 b \bar{b}}|$ which is relatively suppressed. The best point for the excesses predicts $C_{h_1 V V} = -0.36$, $C_{h_1 g g} = - 0.38$, $C_{h_1 \gamma \gamma} = -0.41$, $C_{h_1 t \bar{t}} = -0.37$ and $C_{h_1 b \bar{b}} = -0.18$, and consequently ${\rm Br}(h_1 \to b \bar{b}) = 65\%$ and ${\rm Br}(h_1 \to \gamma \gamma) = 0.6\%$. The pattern ${\rm Br}(h_1 \to b \bar{b})/{\rm Br}_{\rm SM} (H \to b \bar{b}) < 1 $ and  ${\rm Br}(h_1 \to \gamma \gamma )/{\rm Br}_{\rm SM} (H \to \gamma \gamma) \simeq 3.7$ agrees well with the expectation in section II. Moreover, a closer analysis of the samples reveals that the explanations are distributed in three isolated parameter regions
\begin{itemize}
\item {\bf Region I}: $0.06 \lesssim \lambda \lesssim 0.37$,  $0.03 \lesssim \kappa \lesssim 0.17$, $4 \lesssim \tan\beta \lesssim 20 $, $100~{\rm GeV} \lesssim \mu \lesssim 350~{\rm GeV}$ and $\mu/\lambda \sim 800~{\rm GeV}$,
\item {\bf Region II}: $0.22 \lesssim \lambda \lesssim 0.7$,  $0.06 \lesssim \kappa \lesssim 0.6$, $4 \lesssim \tan\beta \lesssim 20 $, $100~{\rm GeV} \lesssim \mu \lesssim 300~{\rm GeV}$ and $\mu/\lambda \sim 250~{\rm GeV}$,
\item {\bf Region III}: $0.37 \lesssim \lambda \lesssim 0.6$,  $0.02 \lesssim \kappa \lesssim 0.14$, $2 \lesssim \tan\beta \lesssim 5 $ and $250~{\rm GeV} \lesssim \mu \lesssim 560~{\rm GeV}$,
\end{itemize}
and they are characterized by
\begin{itemize}
\item The posterior probabilities of the three regions are 0.80, 0.16 and 0.04 respectively\footnote{The concept of the posterior probability comes from Bayesian theorem, which was briefly introduced in~\cite{Fowlie:2016hew}.}. This reflects the fact that the {\bf Region I} is more likely to explain the excesses.
\item In both the {\bf Region I} and the {\bf Region II}, the lightest neutralino $\tilde{\chi}_1^0$ may be either Higgsino-dominated or Singlino-dominated (corresponding to $2 \kappa/\lambda > 1$ and $2 \kappa/\lambda < 1$ respectively~\cite{Ellwanger:2009dp}), while in the {\bf Region III}, $\tilde{\chi}_1^0$ is only Singlino dominated.
\item All the regions are able to predict the central value of the excesses. In the {\bf Region I} and the {\bf Region III}, the most favored parameter points predict $\chi^2_{h_2, {\rm couplings}} \sim 7$, while those in the {\bf Region II} usually predict $\chi^2_{h_2, {\rm couplings}} > 10$. This fact reflects that there is minor tension between the excesses and the data of the discovered Higgs for the {\bf Region II}.
\item The {\bf Region I} and the {\bf Region II} correspond to the lower and upper branches of the first panel in Fig.\ref{fig2}, respectively. For both the branches, $V_{12}$ in Eq.(\ref{Couplings-approximation}) is always negative, and $C_{h_1 b \bar{b}}$ may be either negative (the lower branch) or positive (the upper branch) due to the moderate/strong cancellation between $V_{11} \tan \beta$ and $V_{12}$ in the expression of $C_{h_1 b \bar{b}}$. Similar conclusion applies to the other panels in Fig.\ref{fig2}.
\item For all the regions, $m_{H^\pm} \gtrsim 550~{\rm GeV}$ which is consistent with the results in~\cite{Choi:2019yrv} for a general NMSSM, and $A_1$ may be lighter than 100~{\rm GeV}.
\end{itemize}

We also study the couplings of the SM-like Higgs boson in Fig.\ref{fig3}. This figure shows that the normalized couplings $C_{h_2 V V}$, $C_{h_2 \gamma \gamma}$, $C_{h_2 g g}$ and
$C_{h_2 t \bar{t}}$ are centered around 0.92, and $C_{h_2 b \bar{b}}$ may reach 1.2. ${\rm Br}(h_2 \to b \bar{b})$ varies from $0.55$ to $0.75$ in comparison with its SM prediction $0.575 \pm 0.018$, and ${\rm Br}(h_2 \to \gamma \gamma)$ changes from $1.3 \times 10^{-3}$ to $2.2 \times 10^{-3}$ with its SM prediction $(2.28 \pm 0.11) \times 10^{-3}$~\cite{Heinemeyer:2013tqa}. As pointed out in~\cite{Biekotter:2019kde}, the sizable deviation of the couplings from its SM predictions and the presence of $h_1$ can be explored by future high luminosity LHC or $e^+ e^-$ colliders.

\section{Constraints from DM physics and sparticle search}  \label{Section-constraint}

So far we do not consider the constraints from DM physics and the LHC search for sparticles on the regions. For each sample obtained from the scan in last section, these constraints can be implemented by following procedures~\cite{Cao:2018iyk}:
\begin{itemize}
\item Vary the parameters $\lambda_{\nu}$, $A_{\lambda_{\nu}}$ and $m_{\tilde{\nu}}$ in the sneutrino sector\footnote{Since the soft breaking parameter $A_\nu$ is always associated with the Yukawa coupling $Y_\nu$ (see the term $Y_{\nu} A_{\nu} \tilde{\nu}_R^{\star} \tilde{l} H_u$ in Eq.~\ref{superpotential}) and $Y_\nu \sim 10^{-6}$, the results in this section are insensitive to the value of $A_\nu$ when $|A_\nu|$ is less than several TeV. So we fix $A_{\nu} = 2~{\rm TeV}$ throughout this work.}, and select the sample for which the right-handed sneutrino with $\tau$ flavour is lighter than the other sparticles. To be consistent with the Higgs data fit in the previous section,
 the decay channels $h_1 \to \nu_h \bar{\nu}_h$, $h_2 \to \nu_h \bar{\nu}_h$, $h_1 \to \tilde{\nu}_1 \tilde{\nu}_1$ and $h_2 \to \tilde{\nu}_1\tilde{\nu}_1$ ($\nu_h$ denotes a heavy neutrino with the field $\nu$ as its dominant component) are kinematically forbidden. Since $m_{\nu_h} = 2\lambda_\nu \mu / \lambda$, these requirements are equivalent to
\begin{equation*}
	\lambda_\nu \geq \lambda\times \frac{m_{h_2}}{4\mu},\quad
	m_{\tilde{\nu}_1} \geq \frac{m_{h_2}}{2}.
\end{equation*}
\item Take the sneutrino as the only DM candidate, calculate the quantities such as DM relic density, its scattering rate with nucleon and the photon spectrum of its annihilation in dwarf galaxies, and compare them with relevant measurements of the Planck experiment, the XENON-1T experiment and the Fermion-LAT experiment, respectively.
\item Study the signals of electroweakino production processes at the LHC, and check by simulations whether the signals coincide with the LHC results.
\end{itemize}
Since the involved calculations are rather complex and meanwhile more than 0.1 million samples were accumulated in the scan, it is very time consuming to check all the samples with the constraints. Instead, we only consider one benchmark setting for each of the three regions and illustrate its underlying physics.

\begin{table}[t]
\centering
\begin{tabular}{cc|cc|cc|cc}
\hline
$\lambda$       & 0.164  & $\chi^2_{h_2, {\rm coupling}}$ & 7.97 & $m_{h_1}$                	& 95.9   & ${\rm Br}(h_1 \to \gamma\gamma)$ & $4.88\times10^{-3}$ \\
$\kappa$        & 0.112  & $C_{h_2 Z Z}$                     & 0.918 & $m_{h_2}$                	& 124.6  & ${\rm Br}(h_1 \to b\bar{b})$     & $0.626$ \\
$\tan{\beta}$   & 19.24  & $C_{h_2 W W}$                     & 0.918 & $m_{h_3}$                	& 2332.9 & $C_{h_1 gg}$                     & 0.4152 \\
$\mu$           & 147.7  & $C_{h_2 b \bar{b}}$                     & 0.999 & $m_{A_1}$                	& 301.8  & $C_{h_1 VV}$                     & 0.398 \\
$A_{\lambda}$   & 1785.1 & $C_{h_2 t \bar{t}}$                     & 0.917 & $m_{A_2}$                	& 2332.8 & $V_{11}$                         & 0.0115 \\
$A_{\kappa}$    & -304.6 & $C_{h_2 \tau \bar{\tau}}$                & 0.999 & $m_{H^\pm}$              	& 2348.9 & $V_{12}$                         & 0.3982 \\
$A_{t}$         & 1354.7 & $C_{h_2\gamma \gamma}$              & 0.907 & $m_{\widetilde{\chi}_1^0}$   & 145.1  & $V_{13}$                         & -0.9172 \\
$\mu_{ggF}^{\gamma \gamma}$ & 0.735 & $C_{h_2 g g}$                   & 0.919 & $m_{\widetilde{\chi}_2^0}$   & 155.8  &   $\mu_{\rm CMS}$            & 0.588 \\
 $\mu_{ggF}^{ZZ}$  &  0.753   & $\mu_{VH}^{b\bar{b}}$ &   0.893     & $m_{\widetilde{\chi}_1^\pm}$ & 152.9  &    $\mu_{\rm LEP}$      & 0.119 \\ \hline
\end{tabular}
\caption{{\label{tab:bp1}}Benchmark point of {\bf Region I} with dimensional parameters in unit of GeV. Note that the normalized coupling $C_{h_2 i i^\ast}$ ($i=Z, W, b, t, \tau, \gamma, g$) is equivalent to $\kappa_i$ defined in Table 36 of~\cite{Heinemeyer:2013tqa},
$\mu_{ggF}^{\gamma \gamma}$ denotes the normalized signal strength of $h_2$ for diphoton
decay channel in the gluon fusion production mode, and $\mu_{ggF}^{ZZ}$  and $\mu_{VH}^{b\bar{b}}$ have similar definition to $\mu_{ggF}^{\gamma \gamma}$. Since $C_{h_2 Z Z} \equiv C_{h_2 W W} $ in the theory, $\mu_{ggF}^{WW}= \mu_{ggF}^{ZZ} $~\cite{Heinemeyer:2013tqa}.
This table shows that, in order to explain the excesses, the normalized couplings of $h_2$ are around 0.9 and the signal strengths range from 0.7 to 0.9. This fact reflects a moderate tension between the excesses and the data of the discovered Higgs boson. }
\end{table}

\begin{figure*}[t]
		\centering
        \resizebox{1.0\textwidth}{!}{
        \includegraphics[width=1.0\textwidth]{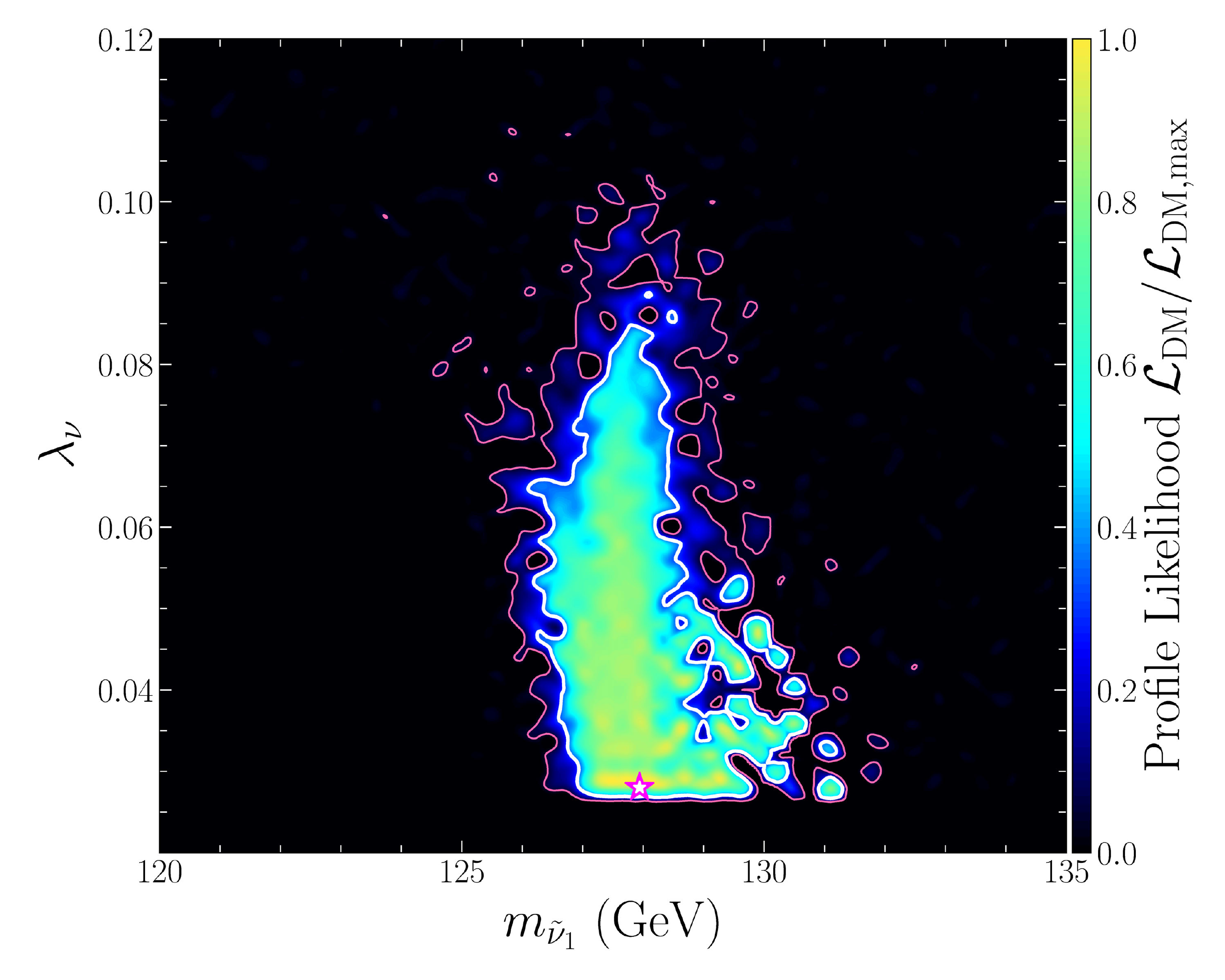}
        \includegraphics[width=1.0\textwidth]{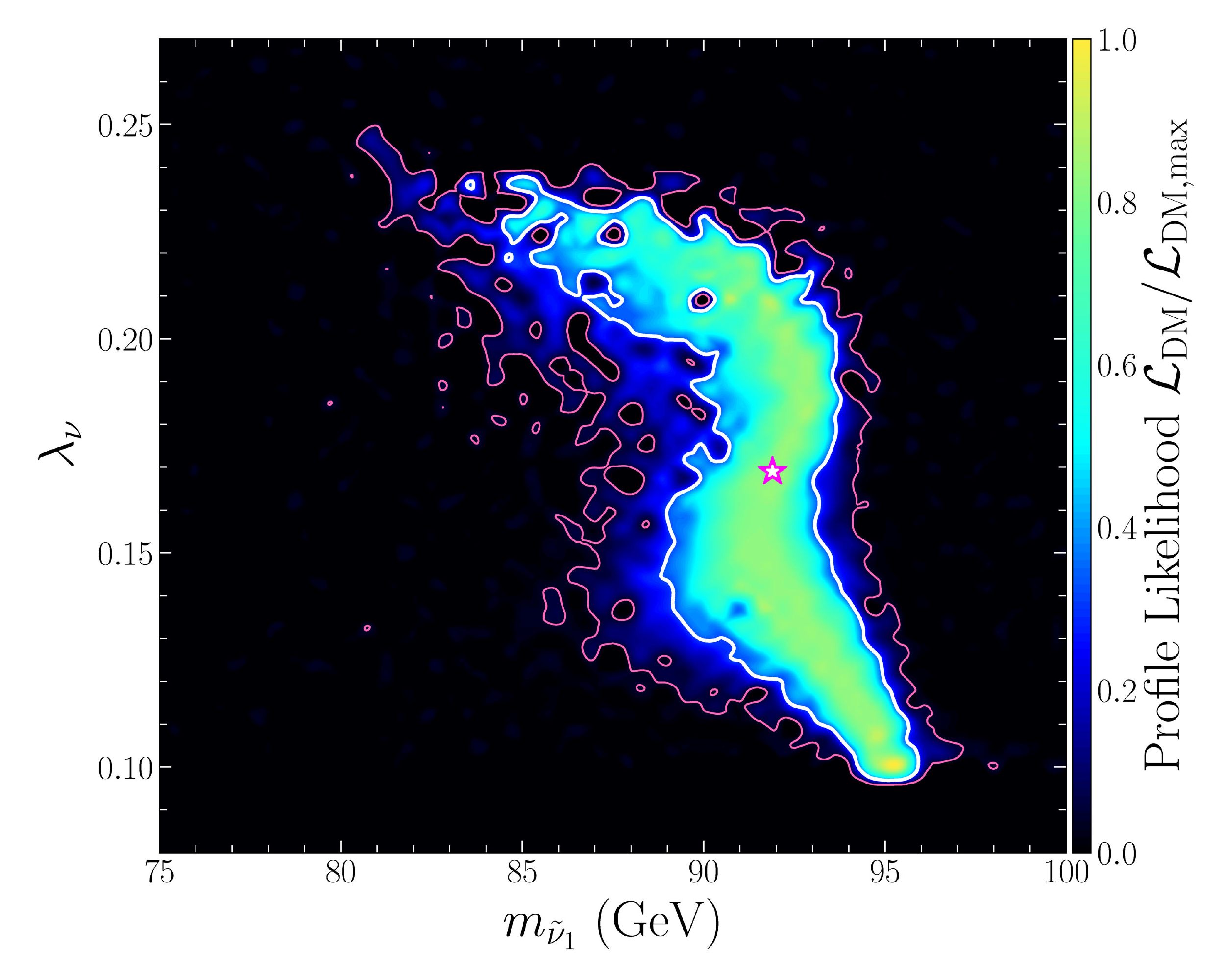}
        }
       \vspace{-0.6cm}
        \caption{The map for the profile likelihood of $\mathcal{L}_{\rm DM} $ in Eq.(\ref{DM-profile}), which is plotted on $\lambda_\nu-m_{\tilde{\nu}_1}$ plane. Given that $\chi^2_{\rm DM, min} \simeq 0$ for the best point which is marked by star symbol,
        the $1 \sigma$ boundary (white solid line) and the $2\sigma$ boundary (red line) correspond to $\chi^2_{\rm DM} \simeq 2.3$ and $\chi^2_{\rm DM} \simeq 6.18$, respectively. The left panel is for the setting of the {\bf Region I}, and the right panel is for the setting of the {\bf Region II}. \label{fig4} }
\end{figure*}	

Let's first consider the benchmark setting of the {\bf Region I}, whose information is presented in Table \ref{tab:bp1}.
We perform a further scan over following region
\begin{eqnarray}
	0 < m_\nu < 150~{\rm GeV},\quad 0 < \lambda_\nu < 0.5, \quad |A_{\lambda_\nu} | < 1{\rm TeV},   \label{DM-parameter}
\end{eqnarray}
with the \texttt{MultiNest} algorithm by requiring $m_{\nu_R} > m_{h_1}/2$ and assuming the sneutrino DM to be CP-even. The likelihood function we adopt is composed by
\begin{eqnarray}
	\mathcal{L}_{\rm DM} = \mathcal{L}_{\Omega_{\tilde{\nu}_1}} \times  \mathcal{L}_{\rm DD} \times \mathcal{L}_{\rm ID},   \label{DM-profile}
\end{eqnarray}
where $\mathcal{L}_{\Omega_{\tilde{\nu}_1}}$, $\mathcal{L}_{\rm DD}$ and $\mathcal{L}_{\rm ID}$ account for the relic density, the XENON-1T experiment and the Fermi-LAT observation of dwarf galaxy respectively, and their explicit forms are presented in~\cite{Cao:2018iyk}.

In the left panel of Fig.\ref{fig4}, we present the profile likelihood of the $\mathcal{L}_{\rm DM}$ for the setting in Table \ref{tab:bp1} on $\lambda_\nu-m_{\tilde{\nu}_1}$ plane with $m_{\tilde{\nu}_1}$ denoting the DM mass. This panel shows that the mass of $\tilde{\nu}_1$ is roughly degenerate with the Higgsino mass $\mu$, which implies that the DM gets the right relic density through co-annihilating with the Higgsinos~\cite{Cao:2018iyk}. Given that $\tilde{\chi}_1^0$ and $\tilde{\chi}_2^0$ in this setting decay by $\tilde{\chi}_{1,2}^0 \to \nu_\tau \tilde{\nu}_1$ and thus they correspond to missing momentum at the LHC, the most promising channel to probe the Higgsinos is through  the process $ p p \to \tilde{\chi}_1^\pm \tilde{\chi}_1^\mp \to (\tau^\pm \tilde{\nu}_1) (\tau^\mp \tilde{\nu}_1)$~\cite{Cao:2018iyk}. Obviously, the LHC has no capability to exclude the moderately light Higgsinos since the $\tau$ leptons are soft due to the compressed mass spectrum of $\tilde{\chi}_1^\pm$ and $\tilde{\nu}_1$~\cite{Cao:2018iyk}. The panel also shows that $\lambda_\nu$ is upper bounded by about 0.1, which means that the DM can not annihilate by the channel $\tilde{\nu}_1 \tilde{\nu}_1 \to h_1 h_1$ to get its right relic density (see the formula of the relic density in various simple DM theories~\cite{Chang:2013oia,Berlin:2014tja}). This is mainly due to the constraint from the DM DD experiments, which may be understood as follows: in the seesaw extension of the NMSSM, the $\tilde{\nu}_1$-nucleon scattering proceeds mainly by the $t$-channel exchange of CP-even Higgs bosons, and any large $\tilde{\nu}_1 \tilde{\nu}_1 h_1$ or $\tilde{\nu}_1 \tilde{\nu}_1 h_2$ coupling is dangerous to spoil the XENON-1T bound. For a CP-even $\tilde{\nu}_1$, the involved coupling strength is given by~\cite{Cao:2018iyk}
\begin{eqnarray}
C_{\tilde{\nu}_1\tilde{\nu}_1 h_i}&=&
\frac{\lambda\lambda_{\nu}M_W}{g}(\sin\beta Z_{i1} + \cos\beta Z_{i2}) - \left[
\frac{\sqrt{2}}{\lambda} \left(2\lambda_{\nu}^2 + \kappa\lambda_{\nu}\right) \mu  - \frac{\lambda_{\nu} A_{\lambda_\nu}}{\sqrt{2}}
\right] Z_{i3}, \label{Csnn}
\end{eqnarray}
where $Z_{ij}$ ($i,j=1,2,3$) denote the elements of the matrix to diagonalize the CP-even Higgs mass matrix in the basis (${\rm Re}[H_d^0]$, ${\rm Re}[H_u^0]$, ${\rm Re}[S]$) with their values given in Table \ref{tab:bp1}. With regard to the specific setting, $C_{\tilde{\nu}_1\tilde{\nu}_1 h_1}/\lambda_\nu$ gets a far dominant contribution from the second bracket in Eq.(\ref{Csnn}), and it is quite large (exceeding 200~{\rm GeV}) since $\mu/\lambda \sim 900~{\rm GeV}$ and $A_{\lambda_\nu}$ is negative\footnote{As shown by the sneutrino mass matrix in ~\cite{Cao:2018iyk}, a negative $A_{\lambda_\nu}$ is needed to ensure that a CP-even sneutrino state is lighter than its CP-odd partner.}. The situation of $C_{\tilde{\nu}_1\tilde{\nu}_1 h_2}/\lambda_\nu$ is quite similar since $|Z_{23}|= 0.39$ is not a small number. Then with the mass insertion method, one can estimate the cross section of the scattering by~\cite{Cao:2019qng}
\begin{eqnarray}
\sigma_{\tilde{\nu}_1-p}^{SI} & \propto & \left \{ \sum_{i=1}^3 (a_{u,i} + a_{d,i}) \right \}^2  \nonumber \\
& \propto & \frac{g^2}{16 m_W^2} \times \left \{ \frac{1}{m_{\tilde{\nu}_1}} \left[
\frac{\sqrt{2}}{\lambda} \left(2\lambda_{\nu}^2 + \kappa\lambda_{\nu}\right) \mu  - \frac{\lambda_{\nu} A_{\lambda_\nu}}{\sqrt{2}}
\right] \frac{m_{h_2}^2 - m_{h_1}^2}{m_{h_1}^2 m_{h_2}^2} \right \}^2 Z_{13}^2 (1 - Z_{13} )^2.  \nonumber
\end{eqnarray}
We checked that this formula is a good approximation of the exact cross section in~\cite{Cao:2018iyk} for the parameter setting. So in order to survive the XENON-1T constraint, $\lambda_\nu$ must be upper bounded by about $0.1$ and correspondingly the co-annihilation channel is dominant. This usually predicts the SI cross section varying from $10^{-48}~{\rm cm^2}$ to $10^{-47}~{\rm cm^2}$, but in some rare cases it may be below $10^{-50}~{\rm cm^2}$. We checked that this conclusion also applies to the case with a CP-odd sneutrino DM, where, although $A_{\lambda_\nu}$ may be either positive or negative to get a CP-odd sneutrino DM~\cite{Cao:2018iyk}, its magnitude is limited so that it can not cancel the contribution of the $\sqrt{2} \mu (2 \lambda_\nu^2 + \kappa \lambda_\nu)/\lambda $ term in an efficient way.


\begin{table}[t]
\centering
\begin{tabular}{cc|cc|cc|cc}
\hline
$\lambda$       & 0.355  & $\chi^2_{h_2, {\rm coupling}}$ & 8.73 & $m_{h_1}$                	& 96.1   & ${\rm Br}(h_1 \to \gamma\gamma)$ & $5.42 \times10^{-3}$ \\
$\kappa$        & 0.433  & $C_{h_2 Z Z}$                & 0.947 & $m_{h_2}$                	& 125.1  & ${\rm Br}(h_1 \to b\bar{b})$     & $0.672$ \\
$\tan{\beta}$   & 15.66  & $C_{h_2 W W}$                & 0.947 & $m_{h_3}$                	& 1623.2 & $C_{h_1 gg}$                     & 0.361 \\
$\mu$           & 115.9  & $C_{h_2 b \bar{b}}$                & 1.135 & $m_{A_1}$                	& 453.3  & $C_{h_1 VV}$                     & 0.321 \\
$A_{\lambda}$   & 1319.1 & $C_{h_2 t \bar{t}}$                 & 0.946 & $m_{A_2}$                	& 1622.4 & $V_{11}$                         & 0.0134 \\
$A_{\kappa}$    & -502.1 & $C_{h_2 \tau \bar{\tau}}$               & 1.135 & $m_{H^\pm}$              	& 1617.5 & $V_{12}$                         & -0.3229 \\
$A_{t}$         & 1901.5 & $C_{h_2\gamma \gamma}$            & 0.918 & $m_{\widetilde{\chi}_1^0}$   & 109.0  & $V_{13}$                         & -0.9463 \\
$\mu_{ggF}^{\gamma \gamma}$ & 0.653 & $C_{h_2 g g}$                 & 0.943 & $m_{\widetilde{\chi}_2^0}$   & 126.4  &  $\mu_{\rm CMS}$         & 0.495 \\
$\mu_{ggF}^{ZZ}$  & 0.694 & $\mu_{VH}^{b\bar{b}}$ &  0.999    & $m_{\widetilde{\chi}_1^\pm}$ & 119.7  & $\mu_{\rm LEP}$    & 0.091 \\ \hline
\end{tabular}
\caption{{\label{tab:bp2}}Same as Table~\ref{tab:bp1}, but for the benchmark setting of the {\bf Region II}.}
\end{table}

Next we turn to the setting of the {\bf Region II} in Table \ref{tab:bp2}, which is featured by $\mu/\lambda \simeq 326~{\rm GeV}$ and $\lambda_\nu \gtrsim 0.08 $. Similar to what we did for the {\bf Region I}, we plot the profile likelihood on $\lambda_\nu-m_{\tilde{\nu}_1}$ plane, and show the boundaries of $1 \sigma$ CI (white solid line) and $2\sigma$ CI (red line) on the right panel of Fig.\ref{fig4}.   We find that the samples in the $2\sigma$ CI annihilated mainly by the channel $\tilde{\nu}_1 \tilde{\nu}_1 \to h_1 h_1$ in early universe. This annihilation requires $\lambda_\nu \sim 0.15$ to get the right relic density~\cite{Chang:2013oia,Berlin:2014tja}, and due to the temperature effect, $\tilde{\nu}_1$ may be lighter than $h_1$ in proceeding the annihilation~\cite{Griest:1990kh}.
We also find that the samples predict the scattering cross section ranging from $10^{-51}~{\rm cm^2}$ to $3 \times 10^{-47}~{\rm cm^2}$, and the constraints from current DM DD experiments is relatively weak. Same as the previous setting, the Higgsinos may be probed by the process $ p p \to \tilde{\chi}_1^\pm \tilde{\chi}_1^\mp \to (\tau^\pm \tilde{\nu}_1) (\tau^\mp \tilde{\nu}_1)$. From the simulation results in~\cite{Cao:2018iyk}, the regions of $\mu  \lesssim 170~{\rm GeV}$ and $\mu \gtrsim 280~{\rm GeV}$ can survive the LHC constraints for the DM mass given in the panel.

\begin{table}[t]
\centering
\begin{tabular}{cc|cc|cc|cc}
\hline
$\lambda$       & 0.434  & $\chi^2_{h_2, {\rm coupling}}$ & 7.80  & $m_{h_1}$                	& 96.1   & ${\rm Br}(h_1 \to \gamma\gamma)$ & $5.74\times10^{-3}$ \\
$\kappa$        & 0.091  & $C_{h_2 Z Z}$                  & 0.921 & $m_{h_2}$                	& 125.3  & ${\rm Br}(h_1 \to b\bar{b})$     & $0.579$ \\
$\tan{\beta}$   & 5.06   & $C_{h_2 W W}$                  & 0.921 & $m_{h_3}$                	& 1595.2 & $C_{h_1 gg}$                     & 0.419 \\
$\mu$           & 317.7  & $C_{h_2 b \bar{b}}$                  & 1.010 & $m_{A_1}$                	& 189.2  & $C_{h_1 VV}$                     & 0.389 \\
$A_{\lambda}$   & 1467.1 & $C_{h_2 t \bar{t}}$                  & 0.918 & $m_{A_2}$                	& 1593.0 & $V_{11}$                         & 0.0379 \\
$A_{\kappa}$    & -175.3 & $C_{h_2 \tau \bar{\tau}}$              & 1.010 & $m_{H^\pm}$              	& 1586.7 & $V_{12}$                         & 0.3891 \\
$A_{t}$         & 1967.1 & $C_{h_2\gamma \gamma}$            & 0.909 & $m_{\widetilde{\chi}_1^0}$ & 132.7  & $V_{13}$                         & -0.9204 \\
$\mu_{ggF}^{\gamma \gamma}$ & 0.724 & $C_{h_2 g g}$                 & 0.917 & $m_{\widetilde{\chi}_2^0}$ & 330.3  & $\mu_{\rm CMS}$   & 0.703   \\
 $\mu_{ggF}^{ZZ}$ & 0.743 & $\mu_{VH}^{b\bar{b}}$   &  0.893      & $m_{\widetilde{\chi}_1^\pm}$ & 324.0  &   $\mu_{\rm LEP}$   & 0.114 \\ \hline
\end{tabular}
\caption{{\label{tab:bp3}}Same as Table~\ref{tab:bp1}, but for the benchmark setting of the {\bf Region III}.}
\end{table}

Finally we consider the benchmark setting of the {\bf Region III} in Table \ref{tab:bp3}. Different from the other settings, now $\tilde{\chi}_1^0$ and $\tilde{\chi}_{2,3}^0$ are Singlino- and Higgsino-dominated respectively with their field compositions given by
\begin{eqnarray}
\tilde{\chi}_1^0 & = & 0.006 \tilde{B}^0 - 0.010 \tilde{W}^0 + 0.058 \tilde{H}^0_d - 0.233 \tilde{H}^0_u + 0.971 \tilde{S}^0, \nonumber \\
\tilde{\chi}_2^0 & = & - 0.020 \tilde{B}^0 + 0.038 \tilde{W}^0 - 0.707 \tilde{H}^0_d + 0.676 \tilde{H}^0_u + 0.205 \tilde{S}^0, \nonumber \\
\tilde{\chi}_3^0 & = & -0.010 \tilde{B}^0 + 0.019 \tilde{W}^0 + 0.705 \tilde{H}^0_d + 0.698 \tilde{H}^0_u + 0.125 \tilde{S}^0.
\end{eqnarray}
In this case, $\tilde{\nu}_1$ is unlikely to co-annihilate with $\tilde{\chi}_1^0$ to get the correct density because the couplings of $\tilde{\chi}_1^0$ with SM particles are rather weak, instead it annihilated mainly by the channels $\tilde{\nu}_1 \tilde{\nu}_1 \to h_i h_j$ with $i,j = 1, 2$ to get the density, which require $m_{\tilde{\nu}_1} \gtrsim 96 {\rm GeV}$ and $\lambda_\nu > 0.1$. Since $\lambda_\nu$ and $\mu/\lambda = 732~{\rm GeV}$ are large in comparison with the other settings, $\sigma^{\rm SI}_{\tilde{\nu}_1 - p} \gtrsim 1 \times 10^{-47}~{\rm cm^2}$ for most cases, and consequently this benchmark setting is limited by the XENON-1T bound. The signals of these electroweakinos at the LHC are as follows. Due to the mass spectrum and field composition, $\tilde{\chi}_1^0$ decays by $\tilde{\chi}_1^0 \to \tilde{\nu}_1 \nu_\tau$ and thus corresponds to missing momentum, $\tilde{\chi}_{2,3}$ decay by the channels $\tilde{\chi}_{2,3}^0 \to Z \tilde{\chi}_1^0,  h_1 \tilde{\chi}_1^0, h_2 \tilde{\chi}_1^0,  A_1 \tilde{\chi}_1^0$ with ${\rm Br} (\tilde{\chi}_2^0 \to \tilde{\chi}_1^0 h_2) = 40\% $, ${\rm Br} (\tilde{\chi}_2^0 \to \tilde{\chi}_1^0 Z) = 31.5\% $, ${\rm Br} (\tilde{\chi}_3^0 \to \tilde{\chi}_1^0 h_2) = 12.5\% $ and ${\rm Br} (\tilde{\chi}_3^0 \to \tilde{\chi}_1^0 Z) = 75.9\% $, and the Higgsino-dominated $\tilde{\chi}_1^\pm$ decays into $\tilde{\chi}_1^0 W^\pm$. The cross sections for the electroweakino pair productions are
\begin{eqnarray}
\sigma ( p p \to \tilde{\chi}_1^0 \tilde{\chi}_1^0 ) &\simeq& 8~{\rm fb}, \quad \sigma ( p p \to \tilde{\chi}_1^0 \tilde{\chi}_2^0 ) \simeq 3~{\rm fb}, \quad \sigma ( p p \to \tilde{\chi}_1^0 \tilde{\chi}_3^0 ) \simeq 10~{\rm fb}, \nonumber \\
\sigma ( p p \to \tilde{\chi}_1^0 \tilde{\chi}_1^\pm ) &\simeq& 47~{\rm fb}, \quad \sigma ( p p \to \tilde{\chi}_2^0 \tilde{\chi}_3^0 ) \simeq 37~{\rm fb}, \quad \sigma ( p p \to \tilde{\chi}_1^\pm \tilde{\chi}_1^\mp ) \simeq 38~{\rm fb}, \nonumber \\
\sigma ( p p \to \tilde{\chi}_2^0 \tilde{\chi}_1^\pm ) &\simeq& \sigma ( p p \to \tilde{\chi}_3^0 \tilde{\chi}_1^\pm ) \simeq 62.5~{\rm fb},
\end{eqnarray}
where we have set the collision energy of the LHC at $13~{\rm TeV}$ and used the package MadGraph/MadEvent~\cite{mad-1,mad-2} in the calculation. These results indicate that the largest signal of the $\tilde{\chi}_1^0$ production is $ p p \to \tilde{\chi}_1^0 \tilde{\chi}_1^\pm \to W^\pm + E_{\rm T}^{\rm miss}$ with its cross section about $50~{\rm fb}$.
This rate, however, is much smaller than current upper bound on Mono-W signal search, which was more than $750~{\rm fb}$ for $m_{\tilde{\chi}_1^0} = 132.7~{\rm GeV}$ by ATLAS analysis with an integrated $36.1~{\rm fb^{-1}}$ data~\cite{Aaboud:2018xdl}. The results also indicate that the best way to explore the setting may be
through the process $p p \to \tilde{\chi}_{2,3} \tilde{\chi}^\pm_1 \to Z W 2 \tilde{\chi}_1^0$ by tri-lepton plus $E_T^{\rm miss}$ signal. In fact, we calculate the $R$ value of the signal by the analysis of CMS collaboration with $35.9~{\rm fb^{-1}}$ data~\cite{Sirunyan:2017lae}\footnote{Note that the CMS analysis of the multi-lepton signal with $35.9~{\rm fb^{-1}}$~\cite{Sirunyan:2017lae} is slightly stronger than corresponding ATLAS analysis with $139~{\rm fb^{-1}}$ data~\cite{ATLAS-newanalysis} in limiting the electroweakinos when $\mu \lesssim 300~{\rm GeV}$. }, like what we did in~\cite{Cao:2018rix}. We find $R = 0.68$ which implies that the setting survives the LHC experiment\footnote{$R\equiv s/s_{95}^{\rm obs}$ is the ratio of theoretical prediction of the signal to its experimental observed $95\%$ C.L. upper limit, therefore $R > 1$ indicates that the theoretical prediction contradicts experimental observation. For more details about the calculation of $R$, see our previous works~\cite{Cao:2018rix,Cao:2018iyk}. }. We also consider several other points in the region, and find that the suppression of ${\rm Br}(\tilde{\chi}_{2,3}^0 \to Z \tilde{\chi}_1^0)$ is vital to survive the collider constraint. Since these points predicts $R > 0.5$, they may be detected by future LHC experiment.

\begin{table}
\centering
\renewcommand{\arraystretch}{1.5}
\begin{tabular}{l|c|c|c|c}
  \hline
  \hline
   \multicolumn{2}{c|}{Region}&DM Annihilation & DM DD Constraint & LHC Signal \\
   \hline
   \hline
 \multirow{2}*{Region I} & $\frac{2\kappa}{\lambda}>1$ & $\tilde{\nu}_1 \tilde{H}$ co-annihilation& Weak & soft $2 \tau + E_T^{\rm miss} $\\ \cline{2-5}
&$\frac{2\kappa}{\lambda}<1$ & $\tilde{\nu}_1 \tilde{\nu}_1 \to h_1 h_1$ &  Strong & $W^{(\ast)} Z^{(\ast)} + E_T^{\rm miss} $ \\
   \hline
 \multirow{2}*{Region II} &$\frac{2\kappa}{\lambda}>1$ &  $\tilde{\nu}_1 \tilde{\nu}_1 \to h_1 h_1$ & Weak & $2 \tau + E_T^{\rm miss} $  \\ \cline{2-5}
  &$\frac{2\kappa}{\lambda}<1$& $\tilde{\nu}_1 \tilde{\nu}_1 \to h_1 h_1$ & Weak &  $W^{(\ast)} Z^{(\ast)} + E_T^{\rm miss} $ \\
   \hline
  Region III &$\frac{2\kappa}{\lambda}<1$& $\tilde{\nu}_1 \tilde{\nu}_1 \to h_1 h_1$ & Moderately strong &  $W^{(\ast)} Z^{(\ast)} + E_T^{\rm miss} $ \\
  \hline
  \hline
\end{tabular}
\caption{Summary of the DM physics and the LHC signals for the {\bf Region I, II, III} discussed in the text.}
\label{table5}
\renewcommand{\arraystretch}{1.0}
\end{table}

Before we end this section, we have following comments:
\begin{itemize}
\item All the three benchmark settings can explain well the excesses and meanwhile keep consistent with the constraints from the DM physics and the LHC search for the electroweakinos by choosing appropriate $\lambda_\nu$, $A_{\lambda_\nu}$ and $m_{\tilde{\nu}}$. Especially, the constraints are rather weak for the {\bf Region I} when $2 \kappa/\lambda > 1$ and the co-annihilation is responsible for the relic density.
\item For both the {\bf Region I} and the {\bf Region II}, $2 \kappa/\lambda$ may be less than 1. This situation is quite similar to that of the {\bf Region III} where
$\tilde{\chi}_1^0$ is Singlino dominated and $\lambda_\nu$ must be larger than 0.1 to get right relic density. Then from our previous discussion
    about the {\bf Region I} and the {\bf Region II}, one can infer that the {\bf Region I} has been tightly limited by DM DD experiments, while the {\bf Region II} is still allowed. We checked the correctness of this conclusion. Moreover, the best way to detect the Higgsinos is through the process $p p \to \tilde{\chi}_{2,3} \tilde{\chi}^\pm_1 \to Z^{(\ast)} W^{(\ast)} 2 \tilde{\chi}_1^0$, and the tri-lepton signal is usually
    suppressed due to the open up of the decay channels $\tilde{\chi}_{2,3}^0 \to h_1 \tilde{\chi}_1^0, h_2 \tilde{\chi}_1^0, A_1 \tilde{\chi}_1^0$.  As a result, the samples in these regions may escape the LHC constraint.
\end{itemize}

In Table \ref{table5}, we summarize the DM physics and the LHC signal of the three regions, which may serve as a guideline to pick out good explanations of the excesses. We also choose a benchmark point for the setting in Table \ref{tab:bp1}, which sets $\lambda_\nu = 0.045$, $A_{\lambda_\nu} = -201.8~{\rm GeV} $, $m_{\tilde{\nu}} = 133.7~{\rm GeV}$ and consequently predicts $\Omega h^2 = 0.1243$ and $\sigma^{\rm SI}_{\tilde{\nu}_1 - p} = 1.6 \times 10^{-47}~{\rm cm^2}$. We calculate its theoretical fine tunings in predicting some measurements, and get $\Delta_Z = 6.37$, $\Delta_{m_{h_1}} = 12.95$, $\Delta_{m_{h_2}} = 62.2$, $\Delta_{\Omega h^2} = 20.2$, $\Delta_{\sigma^{\rm SI}_{\tilde{\nu}_1 -p}} = 8.58$, $\Delta_{\mu_{\rm CMS}} = 6.16$ and $\Delta_{\mu_{\rm LEP}} = 20.65$. In the calculation, we adopt the definition of $\Delta_Z$ and $\Delta_{h_i}$ ($i=1,2,3$) from~\cite{Ellwanger:2011mu} and~\cite{Farina:2013fsa}, respectively, with the input parameters defined at the electroweak scale.
As for the last four fine tunings, they are obtained by maximizing the ratio $\partial \ln O/\partial \ln p_i $ over the input parameter $p_i$ in Eq.(\ref{eq:Higgs-range}) and Eq.(\ref{DM-parameter})
with $O$ denoting an observable. These results indicate that the explanation of the excesses in the seesaw extension is quite natural and thus deserves a careful study.

\section{Conclusions} \label{Section-Conclusion}

The discovery of the SM-like Higgs boson at the LHC validates the Higgs mechanism, while the deficiencies in the Higgs sector of the SM imply a more complex structure to account for the EWSB. The long standing $b\bar{b}$ excess at LEP-II and the continuously observed $\gamma \gamma$ excess by CMS collaboration provide potentially useful hints about the EWSB, and thus they deserve a careful study in new physics models.

In this work we show by both analytic formulas and numerical results that the NMSSM with the Type-I seesaw mechanism can naturally predict the central values of the excesses in certain corners of its parameter space, which are categorized into three regions, and the explanations are consistent with the Higgs data of the discovered Higgs boson, $B-$physics and DM physics measurements, the electroweak precision data as well as the LHC search for sparticles.  This great capability of the theory basically comes from the relaxation of the DM DD constraints. Explicitly speaking, the seesaw mechanism augments the NMSSM  by three generations of right-handed neutrino fields, and renders the right-handed sneutrino as a viable DM candidate.  Due to the gauge singlet nature of the DM, its scattering with nucleon is suppressed in most cases to coincide spontaneously with the latest XENON-1T results. Consequently, broad  parameter spaces in the Higgs sector, especially a light Higgsino mass, are resurrected as experimentally allowed, which makes the theory well suited to explain the excesses.

Our results indicate that the scalar responsible for the excesses should contain a sizable component of the SM Higgs field, and its decay branching ratio into $\gamma \gamma$ state is preferred several times larger than corresponding SM prediction to account for the excesses. The latter can be achieved by a moderately suppressed coupling of the scalar with bottom quarks (in comparison with its other Yukawa couplings) and meanwhile a significant enhancement of its coupling with photons (the chargino-mediated loops play a role in such a process). Correspondingly, the couplings of the SM-like Higgs boson deviates from their SM predictions at a level of $10\%$. If the excesses are corroborated in future, these predictions will serve as the criteria to testify the theory by the precise determination of the scalars' property at next generation $e^+ e^-$ colliders~\cite{Biekotter:2019kde}. Our results also indicate that the explanations are distributed in three isolated parameter regions with different features. These regions can be further classified into five cases according to their underlying DM physics and the LHC signal, which are summarized in Table \ref{table5}. The first case in the table is  least constrained by current measurements in DM physics and the sparticle search at the LHC, while the second case has been tightly limited by the XENON-1T experiment. The Higgsinos in  these cases can survive the LHC constraints by any of following mechanism: the compressed mass spectrum of the Higgsinos with the sneutrino DM, heavy Higgsinos, or the suppression of ${\rm Br}(\tilde{\chi}_{2,3}^0 \to Z \tilde{\chi}_1^0)$. We remind that part of the regions will be explored by updated DM DD experiments and the SUSY search at the LHC, and once new exotic signals are discovered, they will provide complementary information about the EWSB. We also remind that the strong constraints of the XENON-1T experiment on the second case may be avoided in the NMSSM with the inverse seesaw mechanism~\cite{Cao:2019qng}.
The DM physics of this extension is somewhat similar to that of the Type-I extension except that it corresponds to a much more complicated sneutrino sector with several additional parameters, and thus predicts more flexible DM physics.

\section*{Acknowledgements}
This work was supported by the National Natural Science Foundation of China (NNSFC) under Grant Nos. 11575053.


\end{document}